\documentclass{LMCS}

\usepackage{amssymb,amsmath,amscd,latexsym,epic,eepic,psfrag,psfig,graphicx}
\usepackage{algorithm}
\usepackage{macro,luca}
\usepackage{subfigure}
\usepackage[all]{xy}
\usepackage{enumerate,hyperref}

\def\doi{5 (2:7) 2009}
\lmcsheading%
{\doi}
{1--25}
{}
{}
{Jun.~27, 2008}
{May\phantom{.~0}4, 2009}
{}   

\begin{document}

\title[Qualitative Logics and Equivalences for Probabilistic Systems]{
Qualitative Logics and Equivalences \\for Probabilistic Systems\rsuper*}

\author[K.~Chatterjee]{Krishnendu Chatterjee\rsuper a}
\address{{\lsuper a}University of California, Baskin School of Engineering, Santa Cruz, USA}	
\email{c\_krish@eecs.berkeley.edu}

\author[L.\~de Alfaro]{Luca de Alfaro\rsuper b}
\address{{\lsuper b}University of California, Baskin School of Engineering, Santa Cruz, USA}	
\email{luca@soe.ucsc.edu}

\author[M.~Faella]{Marco Faella\rsuper c}
\address{{\lsuper c}Universit\`a di Napoli ``Federico II'', Italy}	
\email{mfaella@na.infn.it} 

\author[A.~Legay]{Axel Legay\rsuper d} 
\address{{\lsuper d}Carnegie Mellon University,
  Computer Science Department, Pittsburgh, USA}
\email{alegay@cs.cmu.edu}

\titlecomment{{\lsuper*}A preliminary version of this paper appeared in the
proceedings of the 4th International Conference on the Quantitative
Evaluation of Systems (QEST 2007).}

\begin{abstract}
We investigate logics and equivalence relations that capture the 
{\em qualitative\/} behavior of Markov Decision Processes (MDPs). 
We present Qualitative Randomized $\ctl$ ($\qrctl$): formulas of this
logic can express the fact that certain temporal properties hold over
all paths, or with probability~0 or~1, but they do not distinguish
among intermediate probability values. 
We present a symbolic, polynomial time model-checking algorithm 
for $\qrctl$ on MDPs.

The logic $\qrctl$ induces an equivalence relation over states of an
MDP that we call {\em qualitative equivalence:\/} informally, two
states are qualitatively equivalent if the sets of formulas that hold
with probability~0 or~1 at the two states are the same. 
We show that for finite \emph{alternating} MDPs, where
nondeterministic and probabilistic choices occur in different states,
qualitative equivalence coincides with alternating bisimulation, 
and can thus be computed via efficient partition-refinement algorithms.
On the other hand, in non-alternating MDPs the
equivalence relations cannot be computed via partition-refinement
algorithms, but rather, they require non-local computation. 
Finally, we consider $\qrctl^*$, that extends $\qrctl$ with nested
temporal operators in the same manner in which $\ctl^*$ extends $\ctl$. We
show that $\qrctl$ and $\qrctl^*$ induce the same qualitative
equivalence on alternating MDPs, while on non-alternating MDPs, the
equivalence arising from $\qrctl^*$ can be strictly finer.
We also provide a full characterization of the relation between
qualitative equivalence, bisimulation, and alternating bisimulation,
according to whether the MDPs are finite, and to whether their
transition relations are finitely-branching.
\end{abstract}

\maketitle\vfill

\section{Introduction}

Markov decision processes (MDPs) provide a model for systems exhibiting both
probabilistic and nondeterministic behavior. 
MDPs were originally introduced to model and solve control problems
for stochastic systems: there, nondeterminism represented the
freedom in the choice of control action, while the probabilistic
component of the behavior described the system's response to the
control action \cite{Bertsekas95}. 
MDPs were later adopted as models for concurrent probabilistic systems, 
probabilistic systems operating in
open environments \cite{SegalaT}, and under-specified probabilistic
systems \cite{bda95,luca-thesis}. 

Given an MDP and a property of interest, we can ask two kinds of 
verification questions: {\em quantitative\/} and 
{\em qualitative\/} questions. 
Quantitative questions relate to the numerical value of the
probability with which the property holds in the system; 
qualitative questions ask whether the property holds with
probability~0 or~1. 
Examples of quantitative questions include the computation of the
maximal and minimal probabilities with which the MDP satisfies a
safety, reachability, or in general, $\omega$-regular property
\cite{bda95}; the corresponding qualitative questions asks whether said
properties hold with probability~0 or~1. 

While much recent work on probabilistic verification has focused on
answering quantitative questions, the interest in qualitative
verification questions predates the one in quantitative ones.
Answering qualitative questions about MDPs is useful in a wide range
of applications. 
In the analysis of randomized algorithms, it is natural to require
that the correct behavior arises with probability~1, and not just with
probability at least $p$ for some $p < 1$. 
For instance, when analyzing a randomized embedded scheduler, we are
interested in whether every thread progresses with probability~1
\cite{EMSOFT05}.  
Such a qualitative question is much easier to study, and to justify,
than its quantitative version; indeed, if we asked for a lower bound
$p < 1$ for the probability of progress, the choice of $p$ would need
to be justified by an analysis of how much failure probability is
acceptable in the final system, an analysis that is generally not easy
to accomplish. 
For the same reason, the correctness of randomized distributed
algorithms is often established with respect to qualitative, rather
than quantitative, criteria 
(see, e.g., \cite{PSL00,KNP_PRISM00,Sto02b}). 
Furthermore, since qualitative answers can generally be computed more
efficiently than quantitative ones, they are often used as a useful
pre-processing step. 
For instance, when computing the maximal probability of reaching a
set of target states $T$, it is convenient to first pre-compute the set
of states $T_1 \supseteq T$ that can reach $T$ with probability~1, and
then compute the maximal probability of reaching $T$: this reduces
the number of states where the quantitative question needs to be
answered, and leads to more efficient algorithms
\cite{luca-kwiat2000}. 
Lastly, we remark that qualitative answers, unlike quantitative ones,
are more robust to perturbations in the numerical values of transition
probabilities in the MDP.
Thus, whenever a system can be modeled only within some
approximation, qualitative verification questions yield information
about the system that is more robust with respect to modeling errors,
and in many ways, more basic in nature. 

In this paper, we provide logics for the specification of qualitative
properties of Markov decision processes, along with model-checking
algorithms for such logics, and we study the equivalence relations
arising from such logics. 
Our starting point for the logics is provided by the probabilistic
logics $\pctl$ and $\pctl^*$ \cite{HanJon94,BerkP95,bda95}.
These logics are able to express bounds on the probability of events: 
the logic $\pctl$ is derived from $\ctl$ by adding to its path quantifiers
$\forall$ (``for all paths'') and $\exists$ (``for at least one path'')
a probabilistic quantifier $\pop$. 
For a bound $q \in [0,1]$, an inequality $\someop \in
\set{<,\leq,\geq,>}$, and a path formula $\phi$, the $\pctl$ formula
$\pop_{\someop q} \phi$ holds at a state if the path formula $\phi$
holds from that state with probability $\someop q$. 
The logic $\pctl^*$ is similarly derived from $\ctl^*$. 
In order to obtain logics for qualitative properties, we consider the
subsets of $\pctl$ and $\pctl^*$ where $\forall$, $\exists$ have been
dropped, and where the bound $q$ against which probabilities are
compared can assume only the two values 0,~1. 
We call the resulting logics $\qrctl$ and $\qrctl^*$, for 
{\em Qualitative Randomized\/} $\ctl$ and $\ctl^*$. 

We provide symbolic model-checking algorithms for the logic 
$\qrctl$; these algorithms can be easily extended to $\qrctl^*$, 
since for MDPs the verification of general temporal-logic
properties can be reduced to reachability questions
\cite{CY95,luca-thesis}.
As usual, the model-checking algorithms for $\qrctl$ proceed by
induction on the structure of a formula. 
The cases for some of the operators are known;
for others, we
give new algorithms, completing the picture of the
symbolic algorithms required for $\qrctl$ model checking. 

We then proceed to study the equivalence relations that arise from
$\qrctl$. 
For two states $s$ and $t$ of an MDP,
we write $s \approx^\pos t$ if the states $s, t$ satisfy the same
$\qrctl$ formulas; similarly, $\qrctl^*$ induces the relation 
$\approx^\pos_*$. 
Informally, $s \approx^\pos t$ holds if the set of properties that
hold with probability~0, positive, and~1, at $s$ and $t$ coincide. 
These relations are thus strictly coarser than standard probabilistic
bisimulation \cite{SL94}, which relates states only when the precise
probability values coincide. 
Other works (\cite{DGJP99}) have introduced \emph{distances}
which quantify the difference in the probabilistic behavior of two MDPs.
When the distance between $s$ and $t$ is zero,
$s$ and $t$ are probabilistically bisimilar, and so they are 
also qualitatively bisimilar.
Aside from that, the distance between two states
is in general unrelated to the states being qualitatively equivalent or not.


The appeal of the relations $\approx^\pos$ and $\approx^\pos_*$ lies
in their ability to relate implementations and specifications in a
qualitative way, abstracting away from precise probability values. 
The relations, and their asymmetrical counterparts related to
simulation, are particularly well-suited to the study of refinement
and implementation of randomized algorithms, where the properties to
be preserved are most often probability-1 properties. 
For instance, when implementing a randomized thread scheduler
\cite{EMSOFT05}, the implementation needs to guarantee that each
thread is scheduled infinitely often with probability~1; it is not
important that the implementation realizes exactly the same
probability of scheduling each thread as the specification. 
Our qualitative relations can also be used as a help to analyze
qualitative properties of systems, similarly to how bisimulation
reductions can help in verification. 
Given a system, the relations enable the construction of a minimized,
qualitatively equivalent system, on which all qualitative questions
about the original system can be answered. 
We will show that our qualitative equivalences are computable by
efficient discrete graph-theoretic algorithms that do not refer to
numerical computation.

We distinguish between {\em alternating\/} MDPs, where probabilistic
and nondeterministic choices occur at different states, from the
general case of {\em non-alternating\/} MDPs, where both choices can occur
at the same state. 
Our first result is that on finite, alternating MDPs, the
relation $\approx^\pos$ coincides with alternating bisimulation
\cite{CONCUR98AHKV} on the MDP regarded as a two-player game of
probability vs.\ nondeterminism. 
This result enables the computation of $\approx^\pos$ via the
efficient partition-refinement algorithms developed for alternating
bisimulation. 
We show that the correspondence between $\approx^\pos$ and alternating
bisimulation breaks down both for infinite MDPs, and for
finite, but non-alternating, MDPs. 
Indeed, we show that on non-alternating MDPs, the relation $\approx^\pos$
cannot be computed by any partition-refinement algorithm that
is {\em local,} in the sense that partitions are refined by looking
only at 1-neighbourhoods of states (the classical partition-refinement
algorithms for simulation and bisimulation are local). 
These results are surprising. 
One is tempted to consider alternating and non-alternating MDPs as
equivalent, since a non-alternating MDP can be translated into an
alternating one by splitting its states into multiple alternating
ones.  
The difference between the alternating and non-alternating
models was already noted in \cite{SegalaTurrini05} for strong and weak
``precise'' simulation, and in \cite{SegalaAxioms} for
axiomatizations.  
Our results indicate that the difference between the
alternating and non-alternating model is even more marked for
$\approx^\pos$, which is a local relation on alternating models, and a
non-local relation in non-alternating ones.

More surprises follow when examining the roles of the $\nextop$
(``next'') and $\mathcal U$ (``until'') operators, and the distinction
between $\qrctl$ and $\qrctl^*$. 
For $\ctl$, it is known that the $\nextop$ operator alone suffices to
characterize bisimulation; the $\until$ operator does not add
distinguishing power. 
The same is true for $\qrctl$ on finite, alternating MDPs. 
On the other hand, we show that for non-alternating, or infinite, MDPs,
$\until$ adds distinguishing power to the logic. 
Similarly, the relations induced by $\qrctl$ and $\qrctl^*$ coincide
on finite, alternating MDPs, but $\qrctl^*$ has greater distinguishing
power, and induces thus finer relations, on non-alternating or infinite
MDPs. 

In summary, we establish that on finite, alternating MDPs,
qualitative equivalence can be computed efficiently, and enjoys many
canonical properties. 
We also show that the situation becomes more complex as soon
as infinite or non-alternating MDPs are considered. 
In all cases, we provide sharp boundaries for the classes of MDPs on
which our statements apply, distinguishing also between finitely and
infinitely-branching MDPs. 
Our results also indicate how the distinction between alternating and
non-alternating MDPs, while often overlooked, is in fact of great
importance where the logical properties of the MDPs are concerned.

Our organization of the paper is as follows:
in section~\ref{sec:defs} we present the formal definitions of  
MDPs and the logics $\qrctl^*$ and $\qrctl$.
In section~\ref{secmodcheck} we present a model checking
algorithm for MDPs with the logic $\qrctl$.
In section~\ref{res1} we characterize the equivalence 
relations of MDPs with respect to $\qrctl$.
In section~\ref{res2} we present algorithms to compute
the equivalence relations.
Finally, in section~\ref{res3} we discuss the roles of
the until and wait-for operators in the logics,
and in section~\ref{res4} we consider the role of 
linear-time nesting (i.e., the equivalences for the
logic $\qrctl^*$).

\section{Definitions}\label{sec:defs}
\subsection{Markov Decision Processes} 

A probability distribution on a countable set $X$ is
a function $f : X \mapsto [0,1]$ such that 
$\sum_{x \in X} f(x) = 1$; 
we denote the set of all probability distributions on $X$ by  
$\distr(X)$. 
Given $f \in \distr(X)$, we define $\supp(f) = \set{x \in X \mid f(x) > 0}$ 
to be the support of $f$. 
We consider a fixed set $\AP$ of atomic propositions, which includes
the distinguished proposition $\tUrn$. 
Given a set $S$, we denote $S^+$ (respectively $S^\omega$)
the set of finite (resp. infinite) sequences of elements of $S$.


A \emph{Markov decision process} (MDP) 
$G=(S,\Acts,\mov,\trans, \lab{\cdot})$ consists of the following components: 
\begin{enumerate}[$\bullet$]
\item a countable set of states $S$;
\item a finite set of actions $\Acts$; 
\item an action assignment $\mov: S \mapsto 2^\Acts \setm \emptyset$,
which associates with each state $s \in S$ the set $\mov(s)$ of
actions that can be chosen at $s$; 
\item a transition function $\trans: S \times \Acts \mapsto\distr(S)$,
which associates with each state $s$ and action $a$ a 
next-state probability distribution $\trans(s,a)$; 
\item a labeling function $\lab{\cdot}: S \mapsto 2^\AP$, 
which labels all $s \in S$ with the set $\lab{s}$ of atomic propositions
true at $s$.
\end{enumerate} 
For $s \in S$ and $a \in \mov(s)$, we let 
$\dest(s,a) = \supp(\trans(s,a))$ be the set of
possible destinations when the action $a$ is chosen at the state $s$. 
The MDP $G$ is \emph{finite} if the state space $S$ is finite, and it is
\emph{finitely-branching} if for all $s \in S$ and $a \in \mov(s)$,
the set $\dest(s,a)$ is finite. 
A {\em play\/} or \emph{path} is an infinite sequence 
$\pats= \seq{s_0,s_1,\ldots} \in S^\omega$ of states of the MDP.
For $s \in S$ and $q \in \AP$, we say that $s$ is a $q$-state iff $q
\in \lab{s}$. 
We define an \emph{edge relation} $E = \set{(s,t) \in S \times S \mid
\exists a \in \mov(s) \qdot t \in \dest(s,a)}$; 
for $s \in S$, we let $E(s) = \set{t \mid (s,t) \in E}$. 
An MDP $G$ is a {\em Markov chain\/} if $|\mov(s)| = 1$ for all $s \in S$; 
in this case, for all $s,t \in S$ we write $\trans(s)(t)$ rather than
$\trans(s,a)(t)$ for the unique $a \in \mov(s)$. 

\paragraph{Interpretations}
We interpret an MDP in two distinct ways: as a $\oneh$-player game, and as a
2-player game. 
In the $\oneh$-player interpretation, probabilistic choice is resolved
probabilistically: 
at a state $s \in S$, player~1 chooses an action $a \in \mov(s)$, 
and the MDP moves to the successor state $t \in S$ with probability
$\trans(s,a)(t)$. 
In the 2-player interpretation, we regard probabilistic choice as
adversarial, and we treat the MDP as a game between player~1 and
player~$p$ ($p$ for ``probability''): 
at a state $s$, player~1 chooses an action $a \in
\mov(s)$, and player~$p$ chooses a destination $t \in \dest(s,a)$. 
The $\oneh$-player interpretation is the classical one \cite{Derman}. 
The 2-player interpretation will be used to relate the qualitative
equivalence relations for the MDP with the alternating relations of
\cite{CONCUR98AHKV}, and thereby derive algorithms for computing the
qualitative equivalence relations.

\paragraph{Strategies}
A \emph{player-1 strategy} is a function $\straa: \; S^+ \mapsto
\distr(\Acts)$ that prescribes the probability distribution
$\straa(\fpat)$ over actions to be played, given the past sequence 
$\fpat \in S^+$ of states visited in the play. 
We require that if $a \in \supp(\straa(\fpat\cdot s))$, then
$a \in \mov(s)$ for all $a \in \Acts$, $s \in S$, and $\fpat\in S^*$. 
We denote by $\Straa$ the set of all player-1 strategies. 

A \emph{player-$p$ strategy} is a function $\strab: \; S^+ \times
\Acts \mapsto \distr(S)$. 
The strategy must be such that, for all $s \in S$, $\fpat \in S^*$,
and $a \in \mov(s)$, we have that 
$\supp(\strab(\fpat\cdot s,a)) \subs \supp(\trans(s,a))$. 
Player~$p$ follows the strategy $\strab$ if, whenever player~1 chooses
move $a$ after a history of play $\fpat$, 
she chooses the destination state with
probability distribution $\strab(\fpat,a)$. 
Thus, in the 2-player interpretation, nondeterminism plays first, and
probability second. 
We denote by $\Strab$ the set of all player-$p$ strategies. 

\paragraph{The 2-player interpretation}
In the 2-player interpretation, once a starting state $s \in S$ and two
strategies $\straa \in \Straa$ and $\strab \in \Strab$ have been
chosen, the game is reduced to an ordinary stochastic process, and it
is possible to define the probabilities of \emph{events,} where an
\emph{event} $\cala \subseteq S^\omega$ is a measurable set of paths. 
We denote the probability of event $\cala$, starting from $s \in S$,
under strategies $\straa \in \Straa$ and $\strab \in \Strab$ by
$\Pr_s^{\straa,\strab}(\cala)$: note that the probability of 
events given strategies $\straa$ and $\strab$ do not depend on the 
transition probabilities of the MDP as the strategy $\strab$ can 
chose any probability distribution at each step.
Given $s \in S$ and $\straa \in \Straa$, $\strab \in \Strab$, 
a play $\langle s_0,s_1, \ldots \rangle$ 
is \emph{feasible} if for every $k \in \Nats$, there is 
$a \in \mov(s_k)$ such that 
$\straa(s_0 ,s_1, \ldots,s_k)(a) >0$
and 
$\strab(s_0 ,s_1, \ldots,s_k,a)(s_{k+1}) >0$. 
We denote by $\outcome(s,\straa,\strab) \subseteq S^\omega$ the 
set of feasible plays that start from $s$ given strategies $\straa$ 
and~$\strab$.

\paragraph{The $\oneh$-player interpretation}
In the $\oneh$-player interpretation, we fix for player~$p$ the strategy
$\strab^*$ that chooses the next state with the distribution
prescribed by $\trans$. 
Precisely, for all $\fpat \in
S^*$, $s \in S$, and $a \in \mov(s)$, we let 
$\strab^*(\fpat\cdot s, a) = \trans(s,a)$. 
We then write $\Pr_s^{\straa}(\cala)$ and 
$\outcome(s,\straa)$ instead of 
$\Pr_s^{\straa,\strab^*}(\cala)$ and 
$\outcome(s,\straa,\strab^*)$, respectively, to underline the fact
that these probabilities and set of outcomes are functions only of the
initial state and of the strategy of player~1. 

\paragraph{Alternating MDPs} 
An \emph{alternating MDP} (AMDP) is an MDP $G=(S,\Acts,\mov,\trans,
\lab{\cdot})$ along with a partition $(S_1, S_p)$ of $S$ such that: 
\begin{enumerate}[(1)]
\item If $s \in S_1$, then $\tUrn \in \lab{s}$ and, for all $a \in
\mov(s)$, $|\dest(s,a)| = 1$.
\item If $s \in S_p$, then $\tUrn \not\in \lab{s}$ and $|\mov(s)| = 1$.
\end{enumerate}
The states in $S_1$ are the player-1, or {\em nondeterministic\/}
states, and the states in $S_p$ are the player-$p$, or {\em
probabilistic\/} states. 
The predicate $\tUrn$ ensures that the MDP is \emph{visibly} alternating:
the difference between player-1 and player-$p$ states is obvious to
the players, and we want it to be obvious to the logic too. 
Alternating MDPs can be represented more succinctly (and more
intuitively) by providing, along with the partition $(S_1, S_p)$ of
$S$, the edge relation $E \subs S \times S$, and a probabilistic
transition function $\ttrans: S_p \mapsto \distr(S)$.  The
probabilistic transition function is defined, for $s \in S_p$, $t \in
S$, and $a \in \mov(s)$, by $\ttrans(s)(t) = \trans(s,a)(t)$.  A {\em
non-alternating\/} MDP is a general (alternating or not) MDP.


We represent MDPs by graphs: vertices correspond to nodes, and each
action $a$ from a state $s$ is drawn as a hyperedge from $s$ to
$\dest(s, a)$.

\subsection{Logics}

We consider two logics for the specification of MDP properties. 
The first, $\qrctl^*$, is a logic that captures {\em qualitative\/}
properties of MDPs, and is a qualitative version of $\pctl^*$ 
\cite{HanJon94,BerkP95,bda95}. 
The logic is defined with respect to the classical, $\oneh$-player
semantics of MDPs. 
The second logic, $\atl^*$, is a game logic defined with respect to
the 2-player semantics of MDPs as in \cite{ATL02}. 

\paragraph{Syntax}

The syntax of both logics is given by defining the set of 
\emph{path formulas} ($\phi$) and \emph{state formulas} ($\psi$) via the
following inductive clauses: 
\[
\begin{array}{lrcl}
\text{path formulas: }
       &\phi & ::= & \psi \mid \neg\phi \mid \phi \lor \phi \mid 
	\bigcirc \phi \mid
        \phi \until \phi \mid \phi \wait \phi;\\
\text{state formulas: }
       &\psi & ::= & \true \mid q \mid \neg \psi \mid 
        \psi \lor \psi \mid \PQ(\phi);
\end{array}
\]
where $q \in \AP$ is an atomic proposition, $\true$ is the boolean
constant with value true, 
and $\PQ$ is a {\em path quantifier.}
The operators $\until$, $\wait$ and $\bigcirc$ are 
temporal operators.
The logics $\atl^*$ and $\qrctl^*$ differ in the path quantifiers:
\begin{enumerate}[$\bullet$]

\item  The path quantifiers in $\qrctl^*$ are: 
        $\exists^{\sure}$, $\forall^{\sure}$, 
        $\exists^{\nullo}$, $\forall^{\nullo}$, 
	$\exists^{\almost},\forall^{\almost},\exists^{\pos}$ and
	$\forall^{\pos}$.

\item The path quantifiers in $\atl^*$ are:
        $\plbr{1},\plbr{p},\plbr{1,p},\plbr{\emptyset}$. 

\end{enumerate}	
The fragments $\atl$ of $\atl^*$ and $\qrctl$ of $\qrctl^*$ consist
of formulas where every temporal operator is immediately preceded by a
path quantifier.  
In the following, when we refer to a ``formula'' of a logic, without
specifying whether it is a state or path formula, we always mean a
state formula. 
As usual, we define $\bo \phi$ and $\diam \phi$ to be abbreviations
for $\phi \wait (\no \true)$ and $\true \until \phi$, respectively. 

\paragraph{Semantics}

For a play $\pats=\seq{s_0,s_1,\ldots}$ we denote by $\pats[i]$ the play 
starting from the $i$-th state of $\pats$, 
i.e., $\pats[i]=\seq{s_i,s_{i+1},\ldots}$.
The semantics for the path formulas is defined as follows, for path
formulas $\phi$, $\phi_1$, $\phi_2$: 
\begin{align*}
\pats &\models \phi_1 \vee \phi_2 && \text{ iff } \pats \models \phi_1
                                     \text{ or }  \pats \models \phi_2 \\
\pats &\models \neg\phi           && \text{ iff } \pats \not\models \phi \\
\pats &\models \bigcirc \phi      && \text{ iff } \pats[1] \models \phi \\
\pats &\models \phi_1 \until \phi_2 && \text{ iff } \exists j\in \Nats. \pats[j] \models \phi_2 \text{ and } \forall 0 \leq i < j. \ \pats[i] \models \phi_1 \\
\pats &\models \phi_1 \wait \phi_2 && \text{ iff }
   \big( \forall j\in \Nats. \ \pats[j] \models \phi_1 \big) \text{ or } 
   \exists j\in \Nats. \ \pats[j] \models \phi_2 \text{ and }
   \forall 0 \leq i \leq j. \ \pats[i] \models \phi_1.
\end{align*}
Observe that 
$$
\neg (\psi_1 \until \psi_2 ) = \Box(\neg \psi_2) \vee (\neg \psi_2  \until 
(\neg \psi_1 \und \neg \psi_2) ) = \neg \psi_2 \wait \neg \psi_1.
$$
Finally, we have
\begin{align*}
\pats \models \psi & \text{ iff } s_0 \models \psi. 
\end{align*}
Given a path formula $\phi$ we denote by 
$\sem{\phi}=\set{\pats \mid \pats \models \phi}$ the
set of plays that satisfy $\phi$.
The semantics of the state formulas of $\atl^*$ and $\qrctl^*$ is 
defined as follows, for a state $s$, path formula $\phi$, and state formulas 
$\psi_1$ and $\psi_2$:
\[
\begin{array}{ll}
s \models \true \\
s \models q &\text{iff } q \in \lab{s} \\
s \models \neg \psi_1 &\text{iff } s \not\models \psi_1 \\
s \models \psi_1 \vee \psi_2 &\text{iff } s \models \psi_1 \text{ or } s \models \psi_2 \\[1ex]

s \models \exists^\sure (\phi) & \text{iff } \exists \straa \in \Straa.\ \outcome(s,\straa) \subseteq \sem{\phi} \\
s \models \forall^\sure (\phi) & \text{iff } \forall \straa \in \Straa.\ \outcome(s,\straa) \subseteq \sem{\phi} \\[1ex]
s \models \exists^\almost (\phi) & \text{iff } \exists \straa \in \Straa.\ \Prb_s^{\straa}(\sem{\phi})=1 \\
s \models \forall^\almost (\phi) & \text{iff } \forall \straa \in \Straa.\ \Prb_s^{\straa}(\sem{\phi})=1 \\[1ex]
s \models \exists^\pos (\phi) & \text{iff } \exists \straa \in \Straa.\ \Prb_s^{\straa}(\sem{\phi})>0 \\
s \models \forall^\pos (\phi) & \text{iff } \forall \straa \in \Straa.\ \Prb_s^{\straa}(\sem{\phi})>0 \\[1ex]
s \models \exists^\nullo (\phi) & \text{iff } \exists \straa \in \Straa.\ \outcome(s,\straa) \inters \sem{\phi} \neq \emptyset \\
s \models \forall^\nullo (\phi) & \text{iff } \forall \straa \in \Straa.\ \outcome(s,\straa) \inters \sem{\phi} \neq \emptyset \\[1ex]
s \models \plbr{1} (\phi) & \text{iff } \exists \straa \in \Straa. \forall \strab \in \Strab. 
	\outcome(s,\straa,\strab) \subseteq \sem{\phi} \\
s \models \plbr{p} (\phi) & \text{iff } \exists \strab \in \Strab. \forall \straa \in \Straa. 
	\outcome(s,\straa,\strab) \subseteq \sem{\phi} \\
s \models \plbr{1,p} (\phi) & \text{iff } \exists \straa \in \Straa. \exists \strab \in \Strab. 
	\outcome(s,\straa,\strab) \subseteq \sem{\phi} \\
s \models \plbr{\emptyset} (\phi) & \text{iff } \forall \straa \in \Straa. \forall \strab \in \Strab. 
	\outcome(s,\straa,\strab) \subseteq \sem{\phi}.\\ 
\end{array}
\]
Given an $\atl^*$ or $\qrctl^*$ formula $\phi$ and an MDP 
$G=(S,\Acts,\mov,\trans, \lab{\cdot})$,
we denote by $\sem{\psi}_G =\set{s\in S \mid s \models \phi}$ the
set of states that satisfy the state formula $\psi$, and we omit the subscript $G$ when
obvious from the context. 
For all path formulas $\phi$ of $\qrctl$, the following
dualities hold:
\begin{equation} \label{eq-dual}
\begin{split}
\sem{\exists^\sure \phi}  &= \sem{\neg(\forall^\nullo (\neg \phi))} \\
\sem{\exists^\nullo \phi} &= \sem{\neg(\forall^\sure (\neg \phi))} \\
\sem{\exists^\pos \phi}   &= \sem{\neg(\forall^\almost (\neg \phi))} \\
\sem{\exists^\almost \phi} &= \sem{\neg(\forall^\pos (\neg \phi))}.
\end{split}
\end{equation}
We now present a simple example to illustrate the difference between
the satisfaction of a path formula with probability~1 and for all paths.

\begin{exa}{}
Consider the simple Markov chain shown in 
Figure~\ref{fig-markov-chains}.
Let the propositions true at states $s$ and $t$ be 
$q$ and $r$, respectively. 
Let us consider the starting state as $s$, and  
the formula $\diam r$ (eventually $r$). 
The formula holds at state $s$ with probability~1, since 
the only closed recurrent set of states in the Markov chain 
is the state $t$ (labeled with proposition $r$).
Hence $\diam r$ holds in state $s$ with probability~1.
However, there is a path (namely, $s^\omega$) that violates
the property eventually $r$, but the probability measure 
for the set $\{ s^\omega\}$ of paths is~0.
Thus the state $s$ does not satisfy that all on all paths 
we have eventually $r$, though it satisfies the property 
eventually $r$ with probability~1.
If we consider the property eventually $q$, then for all 
paths starting from $s$ the property holds (hence the 
property also holds with probability~1).
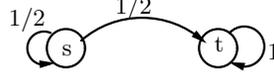
\begin{figure}
   \begin{center}
      \setlength{\unitlength}{0.00043745in}
\begingroup\makeatletter\ifx\SetFigFontNFSS\undefined%
\gdef\SetFigFontNFSS#1#2#3#4#5{%
  \reset@font\fontsize{#1}{#2pt}%
  \fontfamily{#3}\fontseries{#4}\fontshape{#5}%
  \selectfont}%
\fi\endgroup%
{\renewcommand{\dashlinestretch}{30}
\begin{picture}(3090,911)(0,-10)
\thicklines
\put(1590.000,-475.000){\arc{2190.000}{3.9948}{5.4299}}
\blacken\path(2196.695,399.619)(2310.000,350.000)(2233.376,447.101)(2196.695,399.619)
\put(431.250,260.000){\arc{392.946}{1.1584}{5.1248}}
\blacken\path(393.007,39.842)(510.000,80.000)(387.873,99.622)(393.007,39.842)
\put(2800.634,279.142){\arc{476.331}{3.8831}{8.4346}}
\blacken\path(2793.411,71.645)(2670.000,80.000)(2774.960,14.552)(2793.411,71.645)
\put(2490,260){\ellipse{450}{450}}
\put(690,260){\ellipse{450}{450}}
\put(645,170){\makebox(0,0)[lb]{\smash{{\SetFigFontNFSS{9}{10.8}{\rmdefault}{\mddefault}{\updefault}s}}}}
\put(2445,215){\makebox(0,0)[lb]{\smash{{\SetFigFontNFSS{9}{10.8}{\rmdefault}{\mddefault}{\updefault}t}}}}
\put(1275,665){\makebox(0,0)[lb]{\smash{{\SetFigFontNFSS{9}{10.8}{\rmdefault}{\mddefault}{\updefault}$1/2$}}}}
\put(15,530){\makebox(0,0)[lb]{\smash{{\SetFigFontNFSS{9}{10.8}{\rmdefault}{\mddefault}{\updefault}$1/2$}}}}
\put(3075,125){\makebox(0,0)[lb]{\smash{{\SetFigFontNFSS{9}{10.8}{\rmdefault}{\mddefault}{\updefault}$1$}}}}
\end{picture}
}
   \end{center}
 \caption{A simple Markov chain.}\label{fig-markov-chains}  
\end{figure}
\end{exa}

The following lemma establishes a relationship between $\qrctl$ and $\atl$,
proving that the $\qrctl$ quantifiers with superscript
$\sure$ and $\nullo$ are equivalent to the $\atl$ quantifiers.
\begin{lem} \label{lem-equivalence}
For all path formulas $\phi$, the following equivalences hold.
\begin{align*}
\sem{\plbr{1} \phi}         &= \sem{\exists^\sure  \phi} \\
\sem{\plbr{1,p} \phi}       &= \sem{\exists^\nullo \phi} \\
\sem{\plbr{p} \phi}         &= \sem{\forall^\nullo \phi} \\
\sem{\plbr{\emptyset} \phi} &= \sem{\forall^\sure \phi}
\end{align*}
\end{lem}
\begin{proof}
Let $G=(S,\Acts,\mov,\trans, \lab{\cdot})$ be an MDP and let $s\in S$.
We prove the first statement.
Assume $s \models \plbr{1} \phi$.
By definition, there exists $\straa^* \in \Straa$ such that:
$$\forall \strab \in \Strab. 
          \outcome(s,\straa^*,\strab) \subseteq \sem{\phi}.$$
Let $\strab^* \in \Strab$ be the strategy of player~$p$ 
that chooses the next state according to $\trans$
(i.e., the \emph{natural} strategy of player~$p$ in $G$).
We have:
\begin{equation} \label{def-exists-sure}
\outcome(s,\straa^*) = \outcome(s,\straa^*,\strab^*) \subseteq \sem{\phi}.
\end{equation}
Therefore, $s \models \exists^\sure \phi$.

Conversely, assume $s \models \exists^\sure \phi$.
Then, there exists $\straa^* \in \Straa$ such that \eqref{def-exists-sure} holds.
Let $\strab$ be any strategy of player~$p$.
We have that $\outcome(s,\straa^*,\strab) \subseteq \outcome(s,\straa^*,\strab^*)$,
because $\strab^*$ is the most liberal strategy for player~$p$, i.e.,
no player-$p$ strategy can ever choose a successor state that is not 
among those that are chosen by $\strab^*$.
Therefore, $\outcome(s,\straa^*,\strab) \subseteq \sem{\phi}$
and $s \models \plbr{1} \phi$.

Next, we prove the second statement. The remaining statements follow by duality.
Assume $s \models \plbr{1,p} \phi$.
Then, there exist $\straa^\bullet \in \Straa$ 
and $\strab^\bullet \in \Strab$ such that
$\outcome(s,\straa^\bullet,\strab^\bullet) \subseteq \sem{\phi}$.
Let $\strab^*$ be the natural strategy for player~$p$ in $G$.
By the previous argument, $\outcome(s,\straa^\bullet,\strab^\bullet)
\subseteq \outcome(s,\straa^\bullet,\strab^*)$.
Therefore, $\outcome(s,\straa^\bullet,\strab^*)\cap \sem{\phi} \neq \emptyset$
and $s \models \exists^\nullo \phi$. \\
Finally, assume $s \models \exists^\nullo \phi$.
By definition, there exists $\straa^*\in\Straa$ such that
$\outcome(s,\straa^*,\strab^*)\cap \sem{\phi} \neq \emptyset$,
where $\strab^*$ is the natural strategy for player~$p$ in $G$.
Let $\pats$ be a play in $\outcome(s,\straa^*,\strab^*) \cap \sem{\phi}$.
Define $\straa^\bullet$ and $\strab^\bullet$ as the deterministic strategies 
that give as only outcome $\pats$.
We have:
$$\outcome(s,\straa^\bullet,\strab^\bullet) = \set{ \pats } \subseteq \sem{\phi}.$$
Therefore, $s \models \plbr{1,p} \phi$.
\end{proof}
Finally, the following lemma proves the equivalence of some $\qrctl$ formulas.
\begin{lem} \label{lem-trans1}
For all atomic propositions $q, r$, and for all MDPs, we have: 
\begin{align}
  \sem{\exists^\pos \nextop q} &= \sem{\exists^\nullo \nextop q} \notag \\
  \sem{\exists^\almost \nextop q} &= \sem{\exists^\sure \nextop q} \notag \\
  \sem{\exists^\pos q \until r} &= \sem{\exists^\nullo q \until r} \label{eq-transl-pos-until} \\
%
%
  \sem{\exists^\almost q \wait r} &= \sem{\exists^\sure q \wait r}. \notag
%
\end{align}
\end{lem}
\begin{proof}
The first two statements are obvious by definition.
The third statement follows by noting that 
$s \models \exists^\nullo q \until r$ iff there is a finite path in $(S,E)$
from $s$ to an $r$-state, and all states of the path, except possibly
the last, are $q$-states. If such a path exists, there is certainly a strategy
of player~1 that follows it with positive probability.

For the last statement, the ``$\supseteq$'' inclusion is obvious by definition.
For the other inclusion, assume by contradiction that 
$s \in \sem{\exists^\almost q \wait r}$, but all strategies
of player~1 ensuring $q \wait r$ with probability one
also exhibit a path violating it. Then, 
$s \in \sem{\forall^\nullo \: \neg(q \wait r)} = 
       \sem{\forall^\nullo \: \neg r \until \neg q}$.
Following an argument similar to the one for the third statement,
we obtain that 
$s\in \sem{\forall^\pos \: \neg r \until \neg q} = \sem{\forall^\pos \: \neg(q \wait r)}$,
which is a contradiction.
\end{proof}

\subsection{Equivalence Relations}

Given an MDP $G=(S,\Acts,\mov,\trans, \lab{\cdot})$, we consider
the equivalence relations induced over its state space by 
various syntactic subsets of the logics $\qrctl$ and $\atl$.
Define the following fragments of $\qrctl$: 
\begin{enumerate}[$\bullet$] 
\item $\qrctl^\pos$ is the syntactic fragment of $\qrctl$ 
containing only the path quantifiers $\exists^\pos$ and
$\forall^\pos$;
\item $\qrctl^\sure$ is the syntactic fragment of $\qrctl$ 
containing only the path quantifiers $\exists^\sure$ and
$\forall^\sure$.
\end{enumerate}
Note that, because of the dualities
(\ref{eq-dual}), we do not need to consider the fragments for
$\forall^\almost$, $\exists^\almost$, $\forall^\nullo$, $\exists^\nullo$. 
The relations induced by $\qrctl^\pos$ and $\qrctl^\sure$ provide us
with a notion of \emph{qualitative} equivalence between states. 
\begin{align*}
	\approx^\pos & = \set{(s,s') {\in} S {\times} S 
        \mid \forall \psi \in \qrctl^\pos , \  
	s\models \psi \text{ iff } s'\models \psi} \\
	\approx^\sure & = \set{(s,s')  {\in} S {\times} S 
        \mid \forall \psi \in \qrctl^\sure , \ 
        s\models \psi \text{ iff } s'\models \psi}.
\end{align*}
We denote by $\approx^{\pos,\nextop}$ be the equivalence relation defined by
$\qrctl^\pos$, with $\bigcirc$ as the only temporal operator. 
We also define the equivalences $\approx^\pos_*$ and 
$\approx^\sure_*$ as the $\qrctl^*$-version of $\approx^\pos$ and $\approx^\sure$,
respectively.

The syntactic subset of $\atl$ which uses only the path quantifiers
$\plbr{1,p}$ and $\plbr{\emptyset}$ induces the usual notion of
bisimulation \cite{Milner90}: 
indeed, quantifiers $\plbr{1,p}$ and $\plbr{\emptyset}$
correspond to quantifiers $\exists$ and $\forall$ of $\ctl$
\cite{skeletons}, respectively.
The syntactic subset of $\atl$ which uses only the path quantifiers
$\plbr{1}$ and $\plbr{p}$ induces \emph{alternating bisimulation}
\cite{CONCUR98AHKV}.
We have:
\begin{align*}
	\bisim & = \set{(s,s')  \in S \times S 
        \mid \text{ for all $\atl$ formulas } 
	\psi \text{ with } \\
	& \quad \plbr{1,p}, \plbr{\emptyset}
	\text{ as path quantifiers, } s\models \psi \text{ iff } 
	s'\models \psi}; \\[1ex]
	\altsim & = \set{(s,s')  \in S \times S 
        \mid \text{ for all $\atl$ formulas } 
        \psi \text{ with } \\
	& \quad \plbr{1}, \plbr{p}
	\text{ as path quantifiers, }
	s\models \psi \text{ iff } s'\models \psi}; \\[1ex]
	\simclo & = \set{(s,s') \in S \times S \mid \text{ for all
            $\atl$ formulas } \psi, s\models \psi \text{ iff }
          s'\models \psi};
\end{align*}
where $\text{\textrm{TS}}$ is the short form for transition systems.
In the relation $\altsim$, nondeterministic and probabilistic choice
represent the two players of a game. 
In the relation $\bisim$, nondeterminism and probability always cooperate as
a single player.
Finally, the relation $\simclo$ arises from the full logic $\atl$, where
nondeterminism and probability can be either antagonistic or cooperative.
The relations $\bisim$, $\altsim$, and $\simclo$ can be computed 
in polynomial time via well-known partition-refinement algorithms
\cite{Milner90,CONCUR98AHKV}. 

Figure~\ref{turn-results} (resp.\ Figure~\ref{conc-results})
summarizes the relationships between different equivalence relations
on alternating MDPs (resp.\ general MDPs) that we will show in this
paper.  An arrow from relation $A$ to relation $B$ indicates that $A$
implies $B$, i.e., that $A$ is finer than $B$.

\begin{figure}[t]
   \centering
   \setlength{\unitlength}{0.00050000in}
\begingroup\makeatletter\ifx\SetFigFontNFSS\undefined%
\gdef\SetFigFontNFSS#1#2#3#4#5{%
  \reset@font\fontsize{#1}{#2pt}%
  \fontfamily{#3}\fontseries{#4}\fontshape{#5}%
  \selectfont}%
\fi\endgroup%
{\renewcommand{\dashlinestretch}{30}
\begin{picture}(9255,4237)(0,-10)
\blacken\thicklines
\path(7725.000,1726.000)(7665.000,1966.000)(7605.000,1726.000)(7725.000,1726.000)
\path(7665,1966)(7665,466)
\blacken\path(7605.000,706.000)(7665.000,466.000)(7725.000,706.000)(7605.000,706.000)
\blacken\path(8280.000,2251.000)(8040.000,2191.000)(8280.000,2131.000)(8280.000,2251.000)
\path(8040,2191)(9165,2191)
\blacken\path(8925.000,2131.000)(9165.000,2191.000)(8925.000,2251.000)(8925.000,2131.000)
\path(2925,519)(2925,520)(2923,523)
	(2922,527)(2919,534)(2915,544)
	(2910,557)(2904,573)(2897,592)
	(2889,613)(2881,636)(2872,662)
	(2863,689)(2854,718)(2846,748)
	(2837,780)(2829,813)(2821,848)
	(2814,884)(2808,924)(2802,965)
	(2797,1010)(2794,1057)(2791,1108)
	(2790,1161)(2790,1216)(2792,1271)
	(2796,1325)(2800,1377)(2806,1426)
	(2813,1472)(2820,1516)(2828,1557)
	(2837,1596)(2846,1634)(2856,1670)
	(2866,1704)(2876,1737)(2886,1769)
	(2896,1799)(2905,1827)(2915,1853)
	(2923,1877)(2931,1898)(2938,1916)
	(2944,1930)(2948,1942)(2955,1959)
\blacken\path(2919.101,1714.232)(2955.000,1959.000)(2808.139,1759.922)(2919.101,1714.232)
\path(3255,1909)(3256,1908)(3258,1905)
	(3261,1900)(3266,1892)(3274,1882)
	(3283,1868)(3294,1851)(3307,1831)
	(3321,1809)(3336,1784)(3352,1757)
	(3369,1729)(3385,1699)(3401,1667)
	(3417,1634)(3432,1600)(3447,1564)
	(3460,1527)(3473,1487)(3484,1445)
	(3495,1401)(3503,1354)(3509,1304)
	(3514,1252)(3515,1199)(3514,1146)
	(3509,1094)(3503,1045)(3495,999)
	(3484,956)(3473,916)(3460,877)
	(3447,842)(3432,807)(3417,775)
	(3401,744)(3385,714)(3369,686)
	(3352,660)(3336,635)(3321,611)
	(3307,591)(3294,572)(3283,557)
	(3274,544)(3266,534)(3255,519)
\blacken\path(3348.543,748.019)(3255.000,519.000)(3445.312,677.055)(3348.543,748.019)
\blacken\path(7356.795,1733.026)(7440.000,1966.000)(7256.949,1799.590)(7356.795,1733.026)
\path(7440,1966)(7434,1957)(7431,1950)
	(7425,1942)(7419,1931)(7410,1917)
	(7400,1901)(7388,1883)(7375,1862)
	(7359,1839)(7342,1813)(7323,1786)
	(7302,1756)(7280,1725)(7256,1693)
	(7231,1659)(7204,1624)(7175,1589)
	(7145,1552)(7113,1515)(7080,1478)
	(7045,1440)(7008,1402)(6969,1364)
	(6928,1325)(6885,1287)(6839,1248)
	(6791,1208)(6739,1169)(6685,1129)
	(6626,1089)(6564,1048)(6498,1007)
	(6428,966)(6353,925)(6274,884)
	(6191,844)(6105,804)(6015,766)
	(5935,734)(5853,704)(5772,675)
	(5691,648)(5611,623)(5531,600)
	(5454,579)(5377,559)(5302,541)
	(5228,525)(5155,510)(5084,496)
	(5014,484)(4945,472)(4877,462)
	(4810,453)(4744,445)(4679,438)
	(4615,431)(4551,425)(4488,420)
	(4427,415)(4366,411)(4306,407)
	(4248,404)(4191,402)(4136,399)
	(4082,397)(4031,396)(3982,394)
	(3935,393)(3891,393)(3851,392)
	(3813,391)(3778,391)(3747,391)
	(3720,391)(3696,391)(3676,391)
	(3659,391)(3645,391)(3634,391)
	(3627,391)(3615,391)
\blacken\path(3855.000,451.000)(3615.000,391.000)(3855.000,331.000)(3855.000,451.000)
\path(2865,2419)(2863,2419)(2859,2420)
	(2851,2422)(2838,2425)(2820,2429)
	(2797,2435)(2768,2441)(2734,2449)
	(2695,2457)(2651,2467)(2604,2477)
	(2555,2488)(2503,2499)(2451,2510)
	(2398,2521)(2346,2531)(2295,2541)
	(2246,2551)(2198,2560)(2151,2568)
	(2107,2576)(2065,2583)(2024,2589)
	(1986,2595)(1949,2600)(1913,2604)
	(1879,2607)(1846,2610)(1814,2613)
	(1782,2614)(1751,2616)(1720,2616)
	(1690,2616)(1658,2616)(1626,2615)
	(1593,2614)(1561,2612)(1528,2609)
	(1494,2606)(1459,2601)(1423,2596)
	(1385,2591)(1346,2584)(1306,2577)
	(1263,2569)(1219,2560)(1173,2551)
	(1126,2540)(1077,2530)(1027,2518)
	(977,2507)(928,2495)(879,2483)
	(833,2472)(790,2461)(750,2451)
	(715,2442)(686,2434)(662,2428)
	(643,2423)(615,2416)
\blacken\path(833.282,2532.417)(615.000,2416.000)(862.386,2416.000)(833.282,2532.417)
\path(615,2041)(617,2041)(621,2040)
	(629,2038)(642,2036)(660,2033)
	(683,2029)(712,2024)(746,2018)
	(785,2012)(829,2005)(876,1997)
	(925,1989)(977,1980)(1029,1972)
	(1081,1964)(1133,1956)(1184,1948)
	(1234,1940)(1282,1934)(1328,1927)
	(1372,1921)(1415,1916)(1455,1911)
	(1493,1907)(1530,1903)(1566,1900)
	(1600,1897)(1633,1895)(1665,1893)
	(1697,1891)(1728,1890)(1758,1889)
	(1788,1889)(1820,1889)(1852,1889)
	(1884,1890)(1917,1892)(1950,1893)
	(1984,1896)(2018,1899)(2054,1902)
	(2091,1906)(2130,1910)(2170,1915)
	(2212,1921)(2256,1927)(2302,1934)
	(2349,1941)(2397,1949)(2446,1957)
	(2496,1965)(2545,1973)(2593,1982)
	(2639,1990)(2682,1997)(2721,2004)
	(2755,2011)(2785,2016)(2809,2020)
	(2827,2024)(2855,2029)
\blacken\path(2629.285,1927.745)(2855.000,2029.000)(2608.190,2045.876)(2629.285,1927.745)
\path(3565,2029)(3567,2029)(3570,2028)
	(3577,2028)(3587,2027)(3602,2025)
	(3622,2023)(3647,2020)(3678,2017)
	(3714,2013)(3755,2009)(3801,2004)
	(3852,1999)(3906,1993)(3964,1987)
	(4024,1981)(4086,1975)(4149,1969)
	(4213,1963)(4276,1957)(4339,1951)
	(4401,1945)(4462,1940)(4522,1935)
	(4580,1930)(4636,1926)(4690,1922)
	(4742,1918)(4793,1915)(4842,1912)
	(4889,1909)(4935,1907)(4980,1905)
	(5023,1903)(5065,1902)(5107,1901)
	(5147,1900)(5187,1900)(5227,1900)
	(5266,1900)(5305,1901)(5344,1901)
	(5383,1903)(5422,1904)(5462,1906)
	(5501,1908)(5541,1910)(5582,1912)
	(5624,1915)(5666,1919)(5710,1923)
	(5755,1927)(5801,1931)(5848,1936)
	(5898,1942)(5949,1947)(6002,1954)
	(6057,1960)(6113,1968)(6172,1975)
	(6232,1983)(6293,1991)(6356,2000)
	(6420,2009)(6484,2018)(6549,2027)
	(6613,2036)(6675,2046)(6736,2055)
	(6794,2063)(6849,2072)(6901,2079)
	(6947,2086)(6989,2093)(7026,2098)
	(7057,2103)(7082,2107)(7103,2110)
	(7118,2112)(7140,2116)
\blacken\path(6914.604,2014.035)(7140.000,2116.000)(6893.138,2132.100)(6914.604,2014.035)
\blacken\path(7605.000,2581.000)(7665.000,2341.000)(7725.000,2581.000)(7605.000,2581.000)
\path(7665,2341)(7665,3766)
\blacken\path(7725.000,3526.000)(7665.000,3766.000)(7605.000,3526.000)(7725.000,3526.000)
\path(7140,2416)(7138,2416)(7135,2417)
	(7128,2418)(7117,2420)(7101,2423)
	(7080,2427)(7054,2432)(7021,2438)
	(6983,2445)(6940,2453)(6891,2461)
	(6838,2471)(6781,2481)(6720,2491)
	(6657,2502)(6592,2513)(6526,2525)
	(6459,2536)(6393,2547)(6327,2557)
	(6262,2568)(6198,2578)(6136,2587)
	(6075,2596)(6017,2604)(5960,2612)
	(5906,2619)(5853,2626)(5802,2632)
	(5753,2637)(5706,2642)(5660,2647)
	(5615,2650)(5572,2654)(5529,2657)
	(5488,2659)(5447,2661)(5407,2662)
	(5367,2663)(5327,2663)(5288,2663)
	(5248,2663)(5209,2662)(5169,2661)
	(5129,2659)(5089,2657)(5048,2654)
	(5007,2651)(4965,2647)(4922,2643)
	(4877,2638)(4832,2633)(4785,2627)
	(4736,2621)(4686,2614)(4634,2606)
	(4580,2598)(4525,2589)(4467,2579)
	(4409,2569)(4349,2559)(4287,2548)
	(4225,2537)(4162,2525)(4100,2513)
	(4037,2501)(3977,2490)(3917,2478)
	(3861,2467)(3807,2456)(3757,2446)
	(3712,2437)(3671,2429)(3636,2422)
	(3606,2416)(3581,2411)(3561,2407)
	(3547,2404)(3525,2399)
\blacken\path(3745.735,2510.697)(3525.000,2399.000)(3772.329,2393.681)(3745.735,2510.697)
\put(7215,2116){\makebox(0,0)[lb]{\smash{{\SetFigFontNFSS{12}{14.4}{\rmdefault}{\mddefault}{\updefault}$\simclo$}}}}
\put(2640,166){\makebox(0,0)[lb]{\smash{{\SetFigFontNFSS{12}{14.4}{\rmdefault}{\mddefault}{\updefault}$\approx^{\bigcirc,\pos}$}}}}
\put(2865,2116){\makebox(0,0)[lb]{\smash{{\SetFigFontNFSS{12}{14.4}{\rmdefault}{\mddefault}{\updefault}$\approx^\pos$}}}}
\put(7215,3991){\makebox(0,0)[lb]{\smash{{\SetFigFontNFSS{12}{14.4}{\rmdefault}{\mddefault}{\updefault}$\altsim$}}}}
\put(15,2041){\makebox(0,0)[lb]{\smash{{\SetFigFontNFSS{12}{14.4}{\rmdefault}{\mddefault}{\updefault}$\approx^\pos_*$}}}}
\put(1440,2716){\makebox(0,0)[lb]{\smash{{\SetFigFontNFSS{11}{13.2}{\rmdefault}{\mddefault}{\updefault}finite}}}}
\put(5040,2791){\makebox(0,0)[lb]{\smash{{\SetFigFontNFSS{11}{13.2}{\rmdefault}{\mddefault}{\updefault}finite}}}}
\put(4440,1591){\makebox(0,0)[lb]{\smash{{\SetFigFontNFSS{11}{13.2}{\rmdefault}{\mddefault}{\updefault}finite branching}}}}
\put(7215,166){\makebox(0,0)[lb]{\smash{{\SetFigFontNFSS{12}{14.4}{\rmdefault}{\mddefault}{\updefault}$\bisim$}}}}
\put(4440,91){\makebox(0,0)[lb]{\smash{{\SetFigFontNFSS{11}{13.2}{\rmdefault}{\mddefault}{\updefault}finite branching}}}}
\put(2040,1141){\makebox(0,0)[lb]{\smash{{\SetFigFontNFSS{11}{13.2}{\rmdefault}{\mddefault}{\updefault}finite}}}}
\put(9240,2116){\makebox(0,0)[lb]{\smash{{\SetFigFontNFSS{12}{14.4}{\rmdefault}{\mddefault}{\updefault}$\approx^\sure$}}}}
\end{picture}
}
   \caption{Relationship between equivalence relations for AMDPs.}
   \label{turn-results}
\end{figure}
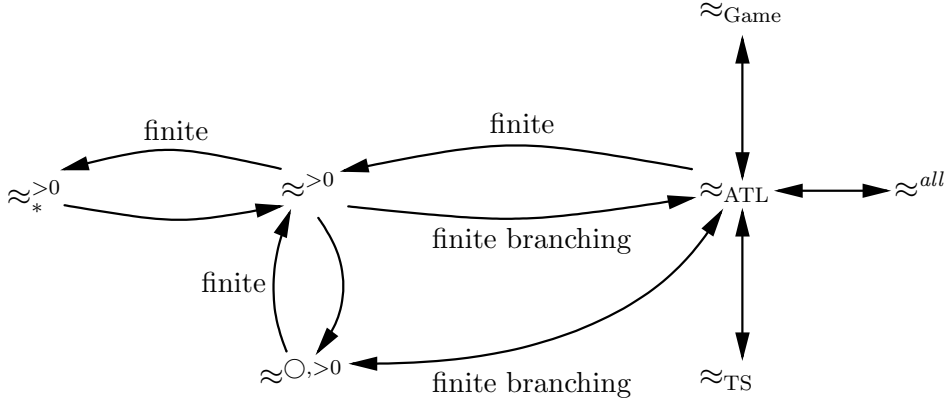

\begin{figure}[t]
   \centering
   \setlength{\unitlength}{0.00050000in}
\begingroup\makeatletter\ifx\SetFigFontNFSS\undefined%
\gdef\SetFigFontNFSS#1#2#3#4#5{%
  \reset@font\fontsize{#1}{#2pt}%
  \fontfamily{#3}\fontseries{#4}\fontshape{#5}%
  \selectfont}%
\fi\endgroup%
{\renewcommand{\dashlinestretch}{30}
\begin{picture}(9180,4255)(0,-10)
\thicklines
\path(7665,1909)(7665,484)
\blacken\path(7605.000,724.000)(7665.000,484.000)(7725.000,724.000)(7605.000,724.000)
\blacken\path(8355.000,2194.000)(8115.000,2134.000)(8355.000,2074.000)(8355.000,2194.000)
\path(8115,2134)(9090,2134)
\blacken\path(8850.000,2074.000)(9090.000,2134.000)(8850.000,2194.000)(8850.000,2074.000)
\path(3240,1909)(3240,1908)(3240,1905)
	(3240,1900)(3240,1892)(3240,1882)
	(3240,1867)(3240,1850)(3241,1830)
	(3241,1806)(3241,1780)(3241,1752)
	(3241,1722)(3242,1690)(3242,1657)
	(3242,1622)(3242,1587)(3243,1550)
	(3243,1512)(3243,1473)(3243,1433)
	(3244,1391)(3244,1347)(3244,1301)
	(3244,1253)(3245,1204)(3245,1153)
	(3245,1102)(3245,1044)(3245,988)
	(3245,936)(3246,889)(3246,846)
	(3246,806)(3246,769)(3246,735)
	(3246,703)(3246,673)(3246,646)
	(3245,619)(3245,595)(3245,573)
	(3245,553)(3245,535)(3245,519)
	(3245,507)(3245,497)(3245,482)
\blacken\path(3185.000,722.000)(3245.000,482.000)(3305.000,722.000)(3185.000,722.000)
\blacken\path(7312.721,1696.868)(7440.000,1909.000)(7227.868,1781.721)(7312.721,1696.868)
\path(7440,1909)(7433,1902)(7429,1897)
	(7422,1890)(7414,1881)(7404,1871)
	(7392,1859)(7378,1844)(7362,1828)
	(7344,1810)(7324,1791)(7302,1769)
	(7278,1747)(7253,1723)(7226,1698)
	(7196,1672)(7166,1645)(7133,1617)
	(7099,1589)(7064,1560)(7026,1530)
	(6986,1500)(6945,1469)(6901,1437)
	(6854,1405)(6805,1372)(6753,1339)
	(6697,1304)(6638,1269)(6575,1232)
	(6508,1195)(6437,1156)(6361,1116)
	(6280,1076)(6195,1035)(6107,993)
	(6015,952)(5937,918)(5858,885)
	(5779,852)(5700,821)(5623,791)
	(5546,762)(5471,735)(5397,709)
	(5325,683)(5253,659)(5184,637)
	(5115,615)(5048,594)(4982,574)
	(4917,554)(4853,536)(4790,518)
	(4727,501)(4666,484)(4605,468)
	(4546,453)(4487,438)(4429,423)
	(4372,409)(4316,396)(4262,383)
	(4210,371)(4159,359)(4110,348)
	(4063,337)(4019,327)(3977,318)
	(3939,310)(3903,302)(3870,295)
	(3841,288)(3815,283)(3792,278)
	(3773,274)(3757,271)(3744,268)
	(3733,266)(3726,264)(3715,262)
\blacken\path(3940.396,363.965)(3715.000,262.000)(3961.862,245.900)(3940.396,363.965)
\path(615,2059)(2790,2059)
\blacken\path(2550.000,1999.000)(2790.000,2059.000)(2550.000,2119.000)(2550.000,1999.000)
\blacken\path(6900.000,1999.000)(7140.000,2059.000)(6900.000,2119.000)(6900.000,1999.000)
\path(7140,2059)(3540,2059)
\path(7665,2359)(7665,3859)
\blacken\path(7725.000,3619.000)(7665.000,3859.000)(7605.000,3619.000)(7725.000,3619.000)
\put(7215,2059){\makebox(0,0)[lb]{\smash{{\SetFigFontNFSS{12}{14.4}{\rmdefault}{\mddefault}{\updefault}$\simclo$}}}}
\put(2640,109){\makebox(0,0)[lb]{\smash{{\SetFigFontNFSS{12}{14.4}{\rmdefault}{\mddefault}{\updefault}$\approx^{\bigcirc,\pos}$}}}}
\put(2865,2059){\makebox(0,0)[lb]{\smash{{\SetFigFontNFSS{12}{14.4}{\rmdefault}{\mddefault}{\updefault}$\approx^\pos$}}}}
\put(7215,4009){\makebox(0,0)[lb]{\smash{{\SetFigFontNFSS{12}{14.4}{\rmdefault}{\mddefault}{\updefault}$\altsim$}}}}
\put(15,2059){\makebox(0,0)[lb]{\smash{{\SetFigFontNFSS{12}{14.4}{\rmdefault}{\mddefault}{\updefault}$\approx^\pos_*$}}}}
\put(4590,1759){\makebox(0,0)[lb]{\smash{{\SetFigFontNFSS{11}{13.2}{\rmdefault}{\mddefault}{\updefault}finite branching}}}}
\put(7215,109){\makebox(0,0)[lb]{\smash{{\SetFigFontNFSS{12}{14.4}{\rmdefault}{\mddefault}{\updefault}$\bisim$}}}}
\put(4515,184){\makebox(0,0)[lb]{\smash{{\SetFigFontNFSS{11}{13.2}{\rmdefault}{\mddefault}{\updefault}finite branching}}}}
\put(9165,2059){\makebox(0,0)[lb]{\smash{{\SetFigFontNFSS{12}{14.4}{\rmdefault}{\mddefault}{\updefault}$\approx^\sure$}}}}
\end{picture}
}
   \caption{Relationship between equivalence relations for MDPs.}
   \label{conc-results}
\end{figure}
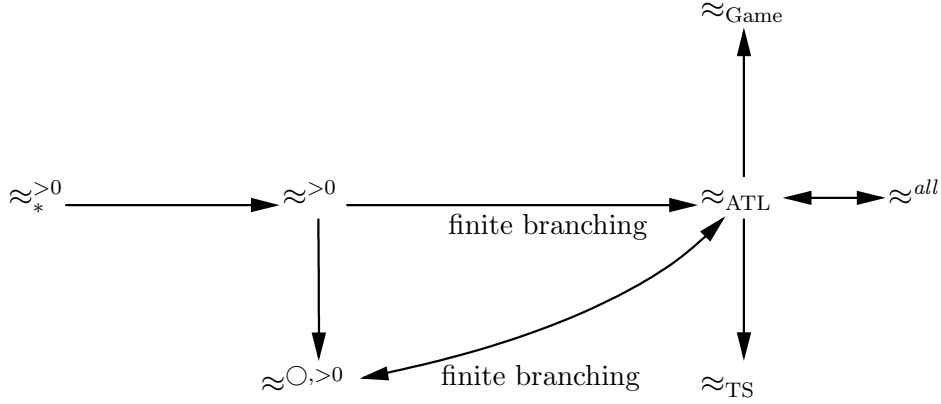

\section{Model Checking \qrctl} 
\label{secmodcheck}
In order to characterize the equivalence relations for $\qrctl$, it is
useful to present first the algorithms for $\qrctl$ model checking. 
The algorithms are based on the results of 
\cite{luca-thesis,luca-stacs97-prob,luca-lics00}; see also
\cite{Trading04}.  
As usual, we present only the algorithms for formulas containing 
one path quantifier, as nested formulas can be model-checked by
recursively iterating the algorithms.
As a consequence of dualities (\ref{eq-dual}), we
need to provide algorithms only for the operators $\exists\nextop$,
$\exists\until$, and $\exists\wait$, and for the modalities $\sure$,
$\almost$, $\pos$, and $\nullo$. 
The algorithms use the following predecessor operators, 
for $X, Y \subs S$: 
\begin{align}
\nonumber 
  \pre(X) & = \set{s \in S \mid \exists a \in \mov(s) \qdot 
    \dest(s,a) \inters X \neq \emptyset} \\[0ex]
\nonumber 
  \cpre(X) & = \set{s \in S \mid \exists a \in \mov(s) \qdot 
    \dest(s,a) \subs X} \\[0ex]
  \begin{split} 
\nonumber 
  \apre(Y,X) & = \set{s \in S \mid \exists a \in \mov(s) \qdot
    \dest(s,a) \subs Y \und \dest(s,a) \inters X \neq \emptyset}.
  \end{split}
\end{align}
The operators $\pre$ and $\cpre$ are classical; the operator $\apre$
is from \cite{luca-focs98}.
We write the algorithms in $\mu$-calculus notation
\cite{Kozen83mu}. 
Given an MDP $G=(S,\Acts,\mov,\trans, \lab{\cdot})$, the interpretation
$\sem{\psi}$ of a $\mu$-calculus formula $\psi$ is a subset of states.
In particular, for a propositional symbol $\ap \in \AP$, we have 
$\sem{\ap} = \set{s \in S \mid \ap \in \lab{s}}$ and 
$\sem{\no\ap} = \set{s \in S \mid \ap \not\in \lab{s}}$. 
The operators $\union$, $\inters$, and the above predecessor operators are
interpreted as the corresponding operations on sets of states,
and $\mu$ and $\nu$ indicate the least and greatest fixpoint, respectively. 
The following result directly leads to model-checking algorithms for
$\qrctl$. 

\begin{thm} \label{theo-mc}
For atomic propositions $q$ and $r$, and for all MDPs, the following
equalities hold:
\begin{align}
%
  \label{sem-next-sure}
  & \sem{\exists^\almost \nextop q} =
    \sem{\exists^\sure \nextop q} = \cpre(\sem{q}) \\[0ex]
  \label{sem-next-pos}
  & \sem{\exists^\pos \nextop q} = 
  \sem{\exists^\nullo \nextop q} = \pre(\sem{q}) \\[0ex]
%
  \label{sem-until-sure} 
  & \sem{\exists^\sure q \until r} 
  = 
  \mu X. ( \sem{r} \union (\sem{q} \inters \cpre(X)) ) \\[0ex]
  \label{sem-until-pos}
  & \sem{\exists^\pos q \until r} = \sem{\exists^\nullo q \until r} = 
  \mu X . ( \sem{r} \union (\sem{q} \inters \pre(X)) ) \\[0ex]
%
  \label{sem-wait-sure}
  & \sem{\exists^\sure q \wait r} =
  \sem{\exists^\almost q \wait r} =
  \nu Y. ( \sem{r} \union ( \sem{q} \inters \cpre(Y)) ) \\[0ex]
  \label{sem-wait-null}
  & \sem{\exists^\nullo q \wait r} =
  \nu Y. ( \sem{r} \union ( \sem{q} \inters \pre(Y)) )
\end{align}
If the MDP is finite, the following equalities also hold:
\begin{align}
%
  \label{sem-until-almost} 
  & \sem{\exists^\almost q \until r} = 
  \nu Y . \mu X . 
  (\sem{r} \union (\sem{q} \inters \apre (Y, X) ) ) \\[0ex]
  \label{sem-wait-pos}
  & \sem{\exists^\pos q \wait r} = 
  \sem{\exists^\pos q \until ((r\wedge q) \vee \exists^\sure \bo q)}.
\end{align}
\end{thm}
\begin{proof}
The formulas involving the $\sure$ and $\nullo$ modalities
(i.e., statements \eqref{sem-next-sure} to \eqref{sem-wait-null})
are derived by the corresponding classical game algorithms, 
thanks to Lemma~\ref{lem-equivalence} and Lemma~\ref{lem-trans1}.
Formula (\ref{sem-until-almost}) is from \cite{luca-focs98}. 
Formula (\ref{sem-wait-pos}) can be understood as follows.  A
\emph{closed component} is a subset of states $T \subs S$ such that,
for all $s \in T$, there is at least one $a \in \mov(s)$ such that
$\dest(s,a) \subs T$.  Using the relation $q \wait r \equiv (q \until
(r\wedge q)) \oder \bo q$ \cite{MPvol1}, we have for $s \in S$ that $s
\models \exists^\pos q \wait r$ iff \emph{(i)}~$s \models \exists^\pos
q \until (q\wedge r)$, or \emph{(ii)}~there is a closed component $T$
composed only of $q$-states, and a path $s_0,s_1, \ldots, s_n$ in
$(S,E)$ composed of $q$-states, with $s_0 = s$ and $s_n \in T$ (see,
e.g., \cite{luca-thesis}). Formula (\ref{sem-wait-pos}) encodes
the disjunction of \emph{(i)} and \emph{(ii)}.
\end{proof}

\noindent
Note that, even though (\ref{sem-wait-pos}) is not a $\mu$-calculus
formula, it can be readily translated into the $\mu$-calculus via 
(\ref{sem-until-pos}) and (\ref{sem-wait-sure}). 
Also observe the $\mu$-calculus formulas corresponding to $\qrctl$
are either alternation free or contain one quantifier alternation
between the $\mu$ and $\nu$ operator.
Thus, from the complexity of evaluating $\mu$-calculus formulas we obtain
the following result.

\begin{thm}\label{theo-mc-complexity}
Given a finite MDP $G=(S,\Acts,\mov,\trans, \lab{\cdot})$ 
and a $\qrctl$ formula $\psi$, the set $\sem{\psi}_G$ can be 
computed in $O(|S| \cdot |\trans| \cdot \ell)$ time, where
$|\trans|=\sum_{s\in S} \sum_{a \in \mov(s)} |\dest(s,a)|$ and 
$\ell$ denotes the length of $\psi$. 
\end{thm}

\begin{proof}
We first consider the computation of $\pre(X)$,
$\cpre(X)$, and $\apre(Y,X)$ for $X,Y\subseteq S$.
To decide whether $s \in \pre(X)$ we check if there exists
$a \in \mov(s)$ such that $\dest(s,a) \cap X \neq \emptyset$.
Similarly, to decide whether $s \in \cpre(X)$ (resp.
$\apre(Y,X)$) we check if there exists $a \in \mov(s)$ such that
$\dest(s,a) \subseteq X$ (resp. $\dest(s,a) \subseteq Y$ and
$\dest(s,a) \cap X \neq \emptyset$).
It follows that given sets $X$ and $Y$, the sets
$\pre(X)$, $\cpre(X)$, and $\apre(Y,X)$ can be computed
in time $O(\sum_{s\in S} \sum_{a \in A} |\dest(s,a)|)$.
Given a formula $\psi$ in $\qrctl$, 
with all of its sub-formulas already evaluated, 
it follows from Theorem~\ref{theo-mc} 
that the computation of $\sem{\psi}$ can be obtained 
by computing a $\mu$-calculus formula of constant length
with at most one quantifier alternation of $\mu$ and $\nu$.
Using the monotonicity property of $\pre,\cpre$ and $\apre$,
and the computation of $\pre,\cpre$ and $\apre$, it follows
that each inner iteration of the $\mu$-calculus formula 
can be computed in 
time $O(\sum_{s\in S} \sum_{a \in A} |\dest(s,a)|)$.
Since the outer iteration of the $\mu$-calculus formula
converges in $|S|$ iterations, it follows that $\semb{\psi}$
can be computed in 
time $O(|S| \cdot \sum_{s\in S} \sum_{a \in A} |\dest(s,a)|)$.
By a bottom-up algorithm that evaluates sub-formulas of a formula
first, we obtain the desired bound for the algorithm.
\end{proof}
\section{Relationship between \qrctl\ and \atl\ Equivalences}
\label{res1}

In this section, we compare the relations induced by \qrctl\ and \atl.
These comparisons will then be used in Section \ref{res2} to derive
algorithms to compute $\approx^\sure$ and $\approx^\pos$.

We first compare $\approx^\sure$ with the relations
induced by \atl.
As a first result, we show that the relations induced by
\atl\ coincide on alternating MDPs (AMDPs).
This result follows from the fact that the turn is visible to the
logic.

\begin{prop} \label{thrm:collapse}
On AMDPs, we have $\altsim \ = \ \bisim$.
\end{prop}
\begin{proof}
Since the turn is observable (via the truth-value of the predicate
$\tUrn$), both $\altsim$ and $\bisim$ can relate only states where the
same player (1 or $p$) can choose the next move. 
Based on this observation, the equality of the relations can be proved
straightforwardly by induction. 
\end{proof}

\begin{cor}\label{coro1}
On AMDPs, we have $\simclo \ = \ \altsim \ =\ \bisim$.
\end{cor}

An immediate consequence of Lemma~\ref{lem-equivalence}
is that $\approx^\sure$ and $\simclo$ coincide. 
This enables the computation of $\approx^\sure$ via the algorithms for
alternating bisimulation \cite{CONCUR98AHKV}. 

\begin{prop}\label{prop:sure}
For all MDPs, $\approx^\sure \ = \ \simclo$.
\end{prop}



Next, we examine the relationship between $\approx^\pos$ and
$\simclo$. 
On finitely-branching MDPs, $\approx^\pos$ is finer than $\simclo$; the
result cannot be extended to infinitely-branching MDPs.

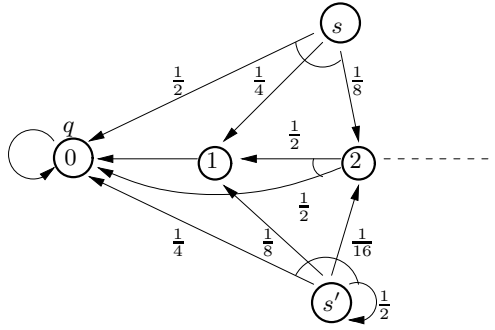
\begin{figure}
   \centering
   \setlength{\unitlength}{0.00062500in}
\begingroup\makeatletter\ifx\SetFigFontNFSS\undefined%
\gdef\SetFigFontNFSS#1#2#3#4#5{%
  \reset@font\fontsize{#1}{#2pt}%
  \fontfamily{#3}\fontseries{#4}\fontshape{#5}%
  \selectfont}%
\fi\endgroup%
{\renewcommand{\dashlinestretch}{30}
\begin{picture}(4150,2714)(0,-10)
\put(2925.500,183.000){\arc{325.000}{4.3176}{8.2488}}
\blacken\path(2983.293,61.801)(2863.000,33.000)(2982.695,1.804)(2983.293,61.801)
\thicklines
\put(2713,178){\ellipse{326}{326}}
\put(2788,2521){\ellipse{326}{326}}
\thinlines
\put(215.500,1420.500){\arc{411.886}{0.5779}{5.7053}}
\blacken\path(314.724,1208.348)(388.000,1308.000)(276.449,1254.555)(314.724,1208.348)
\put(2638.000,2376.750){\arc{451.560}{0.8442}{3.0585}}
\put(2675.500,1345.500){\arc{237.171}{1.8925}{3.4633}}
\put(2662.265,277.853){\arc{562.391}{3.6228}{6.0858}}
\thicklines
\put(1738,1346){\ellipse{272}{272}}
\put(2938,1346){\ellipse{272}{272}}
\put(552,1390){\ellipse{322}{322}}
\thinlines
\path(1588,1383)(763,1383)
\blacken\path(883.000,1413.000)(763.000,1383.000)(883.000,1353.000)(883.000,1413.000)
\path(2788,1383)(1963,1383)
\blacken\path(2083.000,1413.000)(1963.000,1383.000)(2083.000,1353.000)(2083.000,1413.000)
\path(2563,2433)(688,1533)
\blacken\path(783.201,1611.973)(688.000,1533.000)(809.165,1557.882)(783.201,1611.973)
\path(2638,2358)(1813,1533)
\blacken\path(1876.640,1639.066)(1813.000,1533.000)(1919.066,1596.640)(1876.640,1639.066)
\path(2788,2283)(2938,1533)
\blacken\path(2885.049,1644.786)(2938.000,1533.000)(2943.883,1656.553)(2885.049,1644.786)
\path(2563,333)(688,1233)
\blacken\path(809.165,1208.118)(688.000,1233.000)(783.201,1154.027)(809.165,1208.118)
\path(2638,408)(1813,1158)
\blacken\path(1921.973,1099.477)(1813.000,1158.000)(1881.613,1055.081)(1921.973,1099.477)
\path(2713,408)(2938,1158)
\blacken\path(2932.253,1034.440)(2938.000,1158.000)(2874.783,1051.681)(2932.253,1034.440)
\path(2788,1308)(2787,1308)(2785,1307)
	(2781,1305)(2774,1302)(2765,1299)
	(2753,1294)(2738,1288)(2719,1281)
	(2698,1273)(2674,1264)(2647,1254)
	(2618,1243)(2586,1232)(2553,1221)
	(2518,1210)(2482,1198)(2445,1187)
	(2406,1175)(2366,1164)(2324,1154)
	(2281,1143)(2236,1134)(2190,1125)
	(2142,1116)(2091,1108)(2038,1101)
	(1982,1095)(1924,1090)(1863,1086)
	(1801,1084)(1738,1083)(1671,1084)
	(1606,1087)(1543,1091)(1483,1097)
	(1427,1103)(1374,1111)(1324,1120)
	(1276,1129)(1231,1140)(1188,1150)
	(1147,1161)(1108,1173)(1071,1185)
	(1034,1197)(1000,1210)(967,1222)
	(936,1234)(906,1246)(880,1257)
	(855,1267)(834,1276)(815,1284)
	(799,1291)(787,1297)(777,1301)(763,1308)
\blacken\path(883.748,1281.167)(763.000,1308.000)(856.915,1227.502)(883.748,1281.167)
\dashline{60.000}(3153,1383)(4138,1383)
\put(2638,108){\makebox(0,0)[lb]{\smash{{\SetFigFontNFSS{9}{10.8}{\rmdefault}{\mddefault}{\updefault}$s'$}}}}
\put(2713,2433){\makebox(0,0)[lb]{\smash{{\SetFigFontNFSS{9}{10.8}{\rmdefault}{\mddefault}{\updefault}$s$}}}}
\put(463,1608){\makebox(0,0)[lb]{\smash{{\SetFigFontNFSS{9}{10.8}{\rmdefault}{\mddefault}{\updefault}$q$}}}}
\put(477,1328){\makebox(0,0)[lb]{\smash{{\SetFigFontNFSS{9}{10.8}{\rmdefault}{\mddefault}{\updefault}$0$}}}}
\put(2863,1308){\makebox(0,0)[lb]{\smash{{\SetFigFontNFSS{9}{10.8}{\rmdefault}{\mddefault}{\updefault}$2$}}}}
\put(1363,1983){\makebox(0,0)[lb]{\smash{{\SetFigFontNFSS{9}{10.8}{\rmdefault}{\mddefault}{\updefault}$\frac{1}{2}$}}}}
\put(2863,633){\makebox(0,0)[lb]{\smash{{\SetFigFontNFSS{9}{10.8}{\rmdefault}{\mddefault}{\updefault}$\frac{1}{16}$}}}}
\put(1363,633){\makebox(0,0)[lb]{\smash{{\SetFigFontNFSS{9}{10.8}{\rmdefault}{\mddefault}{\updefault}$\frac{1}{4}$}}}}
\put(1663,1308){\makebox(0,0)[lb]{\smash{{\SetFigFontNFSS{9}{10.8}{\rmdefault}{\mddefault}{\updefault}$1$}}}}
\put(2863,1983){\makebox(0,0)[lb]{\smash{{\SetFigFontNFSS{9}{10.8}{\rmdefault}{\mddefault}{\updefault}$\frac{1}{8}$}}}}
\put(2113,633){\makebox(0,0)[lb]{\smash{{\SetFigFontNFSS{9}{10.8}{\rmdefault}{\mddefault}{\updefault}$\frac{1}{8}$}}}}
\put(2038,1983){\makebox(0,0)[lb]{\smash{{\SetFigFontNFSS{9}{10.8}{\rmdefault}{\mddefault}{\updefault}$\frac{1}{4}$}}}}
\put(3088,108){\makebox(0,0)[lb]{\smash{{\SetFigFontNFSS{9}{10.8}{\rmdefault}{\mddefault}{\updefault}$\frac{1}{2}$}}}}
\put(2338,1533){\makebox(0,0)[lb]{\smash{{\SetFigFontNFSS{9}{10.8}{\rmdefault}{\mddefault}{\updefault}$\frac{1}{2}$}}}}
\put(2413,933){\makebox(0,0)[lb]{\smash{{\SetFigFontNFSS{9}{10.8}{\rmdefault}{\mddefault}{\updefault}$\frac{1}{2}$}}}}
\end{picture}
}
   \caption{An infinite Markov chain in which states $s$ and $s'$ cannot 
be distinguished by $\qrctl^{\pos}$, but are distinguished by the $\atl$ 
formula $\plbr{p} \Box \no q$.}
   \label{fig-inf-branch}
\end{figure}

\begin{thm}\label{lemm:possub}
The following assertions hold: 
\begin{enumerate}[\em(1)] 
\item \label{possub-one}
  On finitely-branching MDPs we have 
  $\approx^\pos \ \subseteq \ \simclo$.
\item 
  There is an infinitely-branching AMDP on which 
  $\approx^\pos \ \not\subseteq \ \simclo$.
\end{enumerate}
\end{thm}
\begin{proof}
{\em Assertion 1.} 
For $n > 0$, we consider the $n$-step approximation $\simclo^n$ of
$\simclo$.  
In finite MDPs, we have $\simclo = \simclo^n$ for $n = |S|$;
in finitely-branching MDPs, we have $\simclo = \inters_{n=0}^\infty
\simclo^n$, and this does not extend to MDPs that are not finitely-branching. 
We define a sequence $\Psi_0, \Psi_1, \Psi_2, \ldots$ of sets of formulas
such that, for all $s, t \in S$, we have $s \simclo^n t$ iff $s$ and $t$ 
satisfy the same formulas in $\Psi_n$. 
To this end, given a finite set $\Psi$ of formulas, we denote by
$\clos(\Psi)$ the set of all formulas that consist in disjunctions of
conjunctions of formulas in $\set{\psi, \no \psi \mid \psi \in \Psi}$. 
We assume that each conjunction (resp. disjunction) in $\clos(\Psi)$ does 
not contain repeated elements, so that from the finiteness of $\Psi$
follows the one of $\clos(\Psi)$. 
We let $\Psi_0 = \clos(\AP)$ and, for $k \geq 0$, we let 
$
  \Psi_{k+1} = \clos(\Psi_k \union 
  \set{\exists^\pos \nextop \psi, \exists^\sure \nextop \psi \mid \psi \in \Psi_k})
$.
The formulas in $\clos(\Psi_0),\clos(\Psi_1),\ldots, \clos(\Psi_n)$ 
provide witnesses that $\approx^\pos \ \subseteq \ \simclo^n$.
Thus for all $n$, we have $\approx^\pos \ \subseteq \ \simclo^n$, and
it follows that $\approx^\pos \ \subseteq \ \simclo$.
\smallskip 

\noindent
{\em Assertion 2.} Consider a Markov chain, depicted in Figure~\ref{fig-inf-branch},
with state space $S = \nats \union \set{s,s'}$, with only one predicate symbol $q$, 
such that $\lab{0} = \set{q}$, 
and $\lab{t} = \emptyset$ for all $t \in S \setm \set{0}$. 
There is a transition from $s$ to every $i \in \nats$ 
with probability $1/2^{i+1}$. 
There is a transition from $s'$ to $s'$ with probability $1/2$, and from
$s'$ to every $i \in \nats$ with probability $1/2^{i+2}$. 
There is a transition from $i \in \nats$ with $i > 0$ to every state
in $\set{j \in \nats \mid j < i}$, with uniform probability. 
There is a deterministic transition from $0$ to itself. 
Since this is a Markov chain, the two path quantifiers $\exists$ and
$\forall$ are equivalent, and we need only consider formulas of the
form $\exists^\pos$ and $\exists^\almost$. 
By induction on the length of a $\qrctl$ formula $\phi$, we can then
show that $\phi$ cannot distinguish between states in the set 
$\set{i \in \nats \mid i > |\phi|} \union \set{s,s'}$. 
Hence, 
$s \approx^\pos s'$.  
On the other hand, we have $s \not\simclo s'$, 
since $s \not\models \plbr{p} \bo \no q$ and 
$s' \models \plbr{p} \bo \no q$.
\end{proof}

%

\noindent
To obtain a partial converse of this theorem, we need to translate all $\qrctl$
formulas into $\atl$. 
For finite MDPs, Lemmas~\ref{lem-equivalence} and~\ref{lem-trans1}
enable us to translate all $\qrctl$ formulas, except for formulas of the type
$\exists^\almost \until$ and $\exists^\pos \wait$. 
%
For the latter type, from (\ref{sem-wait-pos}) together
with Lemmas~\ref{lem-equivalence} and~\ref{lem-trans1},
we obtain the following result. 

\begin{lem} \label{lem-trans2}
For finite MDPs, and for all atomic propositions $q, r$, we have 
$$
  \sem{\exists^\pos q \wait r} = \sem{\plbr{1,p}\bigl(q \until
    ((q\wedge r) \oder \plbr{1} \bo q) \bigr)}.
$$
\end{lem}

\noindent
Regarding formulas of the type $\exists^\almost \until$,
they can be model-checked using the $\mu$-calculus
expression (\ref{sem-until-almost}).
To obtain a translation into \atl, 
which will be given in proof of Theorem~\ref{lemm:simsub}, 
we first translate into \atl\ the operator $\apre$. 
To this end, for \atl\ formulas $\phi$, $\psi$, define 
\[
  F_\apre(\phi, \psi) = 
  (\plbr{1} \nextop (\varphi\wedge \psi)) \oder 
  \bigl( \plbr{\emptyset} \nextop \phi \und \plbr{p} \nextop \psi \bigl).
\]

\begin{lem} \label{lem-trans3}  
For AMDPs, and for all $\atl$ formulas $\phi$, $\psi$, we have
$\sem{F_\apre(\phi, \psi)} = \apre(\sem{\phi},\sem{\psi})$.
\end{lem}
\begin{proof}
We consider the following characterization of the $\apre$ 
operator, valid for AMDPs: for sets $X$ and $Y$,
and a state $s$ we have $s \in \apre(Y,X)$ iff the 
following conditions hold:
(a)~if $s \in S_1$, then there exists $a \in \mov(s)$ such
that $\trans(s,a) \in X \cap Y$; and
(b)~if $s \in S_p$, then for the unique action $a \in 
\mov(s)$, we have $\dest(s,a) \subseteq Y$ and
$\dest(s,a) \cap X \neq \emptyset$.
The definition of $F_{\apre}$ captures the above two
conditions. 
The result follows.
\end{proof}

\noindent
Note that the lemma holds only for alternating MDPs: indeed, we will
show that, on non-alternating MDPs, the operator $\apre$ is not
translatable into $\atl$.


Using these lemmas, we can show that on finite AMDPs, we have 
$\simclo \ \subseteq \ \approx^\pos$. 
This result is tight: we cannot relax the assumption that the
MDP is finite, nor the assumption that it is alternating. 

\begin{thm} \label{lemm:simsub}
The following assertions hold: 
\begin{enumerate}[\em(1)]
\item \label{simsub-one} On finite AMDPs, we have $\simclo \ \subseteq \ \approx^\pos$.
\item \label{part-counterexconc} There is a finite MDP on which 
  $\simclo \ \not\subseteq \ \approx^\pos$.  
\item There is an infinite, but finitely-branching, AMDP on which 
  $\simclo \ \not\subseteq \ \approx^\pos$.
\end{enumerate}
\end{thm}

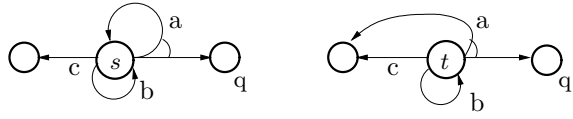
\begin{figure}
   \centering
   \setlength{\unitlength}{0.00041667in}
\begingroup\makeatletter\ifx\SetFigFont\undefined%
\gdef\SetFigFont#1#2#3#4#5{%
  \reset@font\fontsize{#1}{#2pt}%
  \fontfamily{#3}\fontseries{#4}\fontshape{#5}%
  \selectfont}%
\fi\endgroup%
{\renewcommand{\dashlinestretch}{30}
\begin{picture}(7144,1404)(0,-10)
\put(1614.864,1034.318){\arc{690.288}{2.7886}{7.9232}}
\blacken\path(1240.784,1028.041)(1291.000,915.000)(1299.888,1038.373)(1240.784,1028.041)
\put(1891.000,765.000){\arc{335.410}{5.1760}{6.7468}}
\put(5791.000,765.000){\arc{335.410}{5.1760}{6.7468}}
\put(1341.735,431.912){\arc{543.381}{5.8740}{10.1642}}
\blacken\path(1642.153,427.379)(1591.000,540.000)(1583.137,416.557)(1642.153,427.379)
\put(5496.357,341.786){\arc{503.685}{5.7720}{10.3308}}
\blacken\path(5776.434,357.075)(5716.000,465.000)(5718.536,341.333)(5776.434,357.075)
\thicklines
\put(203,690){\ellipse{376}{376}}
\put(6841,653){\ellipse{376}{376}}
\put(1356,653){\ellipse{470}{470}}
\put(2753,690){\ellipse{376}{376}}
\put(5563,661){\ellipse{470}{470}}
\put(4253,690){\ellipse{376}{376}}
\thinlines
\path(1591,690)(2566,690)
\blacken\path(2446.000,660.000)(2566.000,690.000)(2446.000,720.000)(2446.000,660.000)
\path(5791,690)(6616,690)
\blacken\path(6496.000,660.000)(6616.000,690.000)(6496.000,720.000)(6496.000,660.000)
\path(1141,690)(391,690)
\blacken\path(511.000,720.000)(391.000,690.000)(511.000,660.000)(511.000,720.000)
\path(5341,690)(4441,690)
\blacken\path(4561.000,720.000)(4441.000,690.000)(4561.000,660.000)(4561.000,720.000)
\path(5791,690)(5792,691)(5794,694)
	(5797,700)(5802,708)(5809,719)
	(5817,733)(5826,750)(5836,769)
	(5847,791)(5858,814)(5868,838)
	(5878,862)(5886,888)(5892,913)
	(5896,939)(5898,964)(5896,990)
	(5891,1016)(5882,1041)(5867,1067)
	(5848,1092)(5822,1117)(5791,1140)
	(5760,1158)(5726,1173)(5692,1187)
	(5659,1199)(5627,1208)(5597,1216)
	(5569,1222)(5544,1227)(5521,1231)
	(5500,1234)(5481,1237)(5463,1238)
	(5445,1239)(5428,1240)(5411,1240)
	(5393,1241)(5374,1241)(5353,1241)
	(5329,1240)(5303,1240)(5273,1239)
	(5240,1238)(5202,1237)(5161,1235)
	(5116,1232)(5068,1228)(5017,1222)
	(4966,1215)(4908,1204)(4853,1192)
	(4802,1178)(4755,1163)(4712,1147)
	(4674,1131)(4638,1114)(4605,1097)
	(4575,1079)(4546,1061)(4520,1043)
	(4495,1025)(4472,1008)(4451,991)
	(4432,975)(4415,960)(4401,948)
	(4389,937)(4380,928)(4366,915)
\blacken\path(4433.522,1018.638)(4366.000,915.000)(4474.349,974.670)(4433.522,1018.638)
\put(1666,165){\makebox(0,0)[lb]{\smash{{\SetFigFont{10}{12.0}{\rmdefault}{\mddefault}{\updefault}b}}}}
\put(766,465){\makebox(0,0)[lb]{\smash{{\SetFigFont{10}{12.0}{\rmdefault}{\mddefault}{\updefault}c}}}}
\put(5866,15){\makebox(0,0)[lb]{\smash{{\SetFigFont{10}{12.0}{\rmdefault}{\mddefault}{\updefault}b}}}}
\put(2866,315){\makebox(0,0)[lb]{\smash{{\SetFigFont{9}{10.8}{\rmdefault}{\mddefault}{\updefault}q}}}}
\put(2041,1065){\makebox(0,0)[lb]{\smash{{\SetFigFont{10}{12.0}{\rmdefault}{\mddefault}{\updefault}a}}}}
\put(6991,315){\makebox(0,0)[lb]{\smash{{\SetFigFont{9}{10.8}{\rmdefault}{\mddefault}{\updefault}q}}}}
\put(5941,1065){\makebox(0,0)[lb]{\smash{{\SetFigFont{10}{12.0}{\rmdefault}{\mddefault}{\updefault}a}}}}
\put(4816,465){\makebox(0,0)[lb]{\smash{{\SetFigFont{10}{12.0}{\rmdefault}{\mddefault}{\updefault}c}}}}
\put(1291,540){\makebox(0,0)[lb]{\smash{{\SetFigFont{9}{10.8}{\rmdefault}{\mddefault}{\updefault}$s$}}}}
\put(5491,540){\makebox(0,0)[lb]{\smash{{\SetFigFont{9}{10.8}{\rmdefault}{\mddefault}{\updefault}$t$}}}}
\end{picture}
}
   \caption{States $s$ and $t$ cannot be distinguished by $\atl$, 
but are distinguished by $\exists^\almost \Diamond q$.}
   \label{fig4}
\end{figure}

\begin{figure}[t] 
   \centering
   \setlength{\unitlength}{0.00058333in}
\begingroup\makeatletter\ifx\SetFigFont\undefined%
\gdef\SetFigFont#1#2#3#4#5{%
  \reset@font\fontsize{#1}{#2pt}%
  \fontfamily{#3}\fontseries{#4}\fontshape{#5}%
  \selectfont}%
\fi\endgroup%
{\renewcommand{\dashlinestretch}{30}
\begin{picture}(4915,2359)(0,-10)
\put(1453.000,1873.000){\arc{474.342}{5.9614}{7.5322}}
\put(1528.000,523.000){\arc{335.410}{5.1760}{6.7468}}
\put(265.274,1873.000){\arc{451.120}{5.9442}{7.4927}}
\put(2661.366,1873.000){\arc{458.499}{5.9499}{7.4523}}
\put(2665.146,523.000){\arc{313.871}{5.1952}{6.7815}}
\put(334.046,523.000){\arc{324.640}{5.0709}{6.7635}}
\thicklines
\put(1378,1911){\ellipse{272}{272}}
\put(2578,1911){\ellipse{272}{272}}
\put(3778,1911){\ellipse{272}{272}}
\put(192,1955){\ellipse{322}{322}}
\put(1406,431){\ellipse{272}{272}}
\put(2531,431){\ellipse{272}{272}}
\put(3703,411){\ellipse{272}{272}}
\put(178,453){\ellipse{326}{326}}
\put(778,1161){\ellipse{272}{272}}
\put(3103,1161){\ellipse{272}{272}}
\put(1903,1161){\ellipse{272}{272}}
\thinlines
\path(2728,1948)(3628,1948)
\blacken\path(3508.000,1918.000)(3628.000,1948.000)(3508.000,1978.000)(3508.000,1918.000)
\path(1528,1948)(2428,1948)
\blacken\path(2308.000,1918.000)(2428.000,1948.000)(2308.000,1978.000)(2308.000,1918.000)
\path(368,1948)(1228,1948)
\blacken\path(1108.000,1918.000)(1228.000,1948.000)(1108.000,1978.000)(1108.000,1918.000)
\path(253,1798)(665,1235)
\blacken\path(569.923,1314.123)(665.000,1235.000)(618.343,1349.556)(569.923,1314.123)
\dashline{60.000}(3853,1798)(4078,1423)
\dashline{60.000}(3918,1948)(4903,1948)
\path(287,563)(703,1048)
\blacken\path(647.645,937.384)(703.000,1048.000)(602.103,976.447)(647.645,937.384)
\path(328,448)(1228,448)
\blacken\path(1108.000,418.000)(1228.000,448.000)(1108.000,478.000)(1108.000,418.000)
\path(1528,448)(2428,448)
\blacken\path(2308.000,418.000)(2428.000,448.000)(2308.000,478.000)(2308.000,418.000)
\path(2653,448)(3553,448)
\blacken\path(3433.000,418.000)(3553.000,448.000)(3433.000,478.000)(3433.000,418.000)
\dashline{60.000}(3767,535)(4078,898)
\dashline{60.000}(3853,448)(4903,448)
\path(1491,526)(1828,1048)
\blacken\path(1788.118,930.913)(1828.000,1048.000)(1737.710,963.456)(1788.118,930.913)
\path(1439,1788)(1814,1263)
\blacken\path(1719.839,1343.211)(1814.000,1263.000)(1768.663,1378.085)(1719.839,1343.211)
\path(2660,1803)(3035,1278)
\blacken\path(2940.839,1358.211)(3035.000,1278.000)(2989.663,1393.085)(2940.839,1358.211)
\path(2638,526)(3028,1048)
\blacken\path(2980.210,933.912)(3028.000,1048.000)(2932.144,969.823)(2980.210,933.912)
\put(2503,2173){\makebox(0,0)[lb]{\smash{{\SetFigFont{8}{9.6}{\rmdefault}{\mddefault}{\updefault}$q$}}}}
\put(3703,2173){\makebox(0,0)[lb]{\smash{{\SetFigFont{8}{9.6}{\rmdefault}{\mddefault}{\updefault}$q$}}}}
\put(1303,2173){\makebox(0,0)[lb]{\smash{{\SetFigFont{8}{9.6}{\rmdefault}{\mddefault}{\updefault}$q$}}}}
\put(103,2173){\makebox(0,0)[lb]{\smash{{\SetFigFont{8}{9.6}{\rmdefault}{\mddefault}{\updefault}$q$}}}}
\put(3628,73){\makebox(0,0)[lb]{\smash{{\SetFigFont{8}{9.6}{\rmdefault}{\mddefault}{\updefault}$q$}}}}
\put(2428,73){\makebox(0,0)[lb]{\smash{{\SetFigFont{8}{9.6}{\rmdefault}{\mddefault}{\updefault}$q$}}}}
\put(103,73){\makebox(0,0)[lb]{\smash{{\SetFigFont{8}{9.6}{\rmdefault}{\mddefault}{\updefault}$q$}}}}
\put(1303,73){\makebox(0,0)[lb]{\smash{{\SetFigFont{8}{9.6}{\rmdefault}{\mddefault}{\updefault}$q$}}}}
\put(553,1498){\makebox(0,0)[lb]{\smash{{\SetFigFont{8}{9.6}{\rmdefault}{\mddefault}{\updefault}$1-x_1$}}}}
\put(1678,1498){\makebox(0,0)[lb]{\smash{{\SetFigFont{8}{9.6}{\rmdefault}{\mddefault}{\updefault}$1-x_2$}}}}
\put(2900,1536){\makebox(0,0)[lb]{\smash{{\SetFigFont{8}{9.6}{\rmdefault}{\mddefault}{\updefault}$1-x_3$}}}}
\put(117,1893){\makebox(0,0)[lb]{\smash{{\SetFigFont{8}{9.6}{\rmdefault}{\mddefault}{\updefault}$t$}}}}
\put(116,380){\makebox(0,0)[lb]{\smash{{\SetFigFont{8}{9.6}{\rmdefault}{\mddefault}{\updefault}$s$}}}}
\put(628,223){\makebox(0,0)[lb]{\smash{{\SetFigFont{8}{9.6}{\rmdefault}{\mddefault}{\updefault}$y_1$}}}}
\put(1753,223){\makebox(0,0)[lb]{\smash{{\SetFigFont{8}{9.6}{\rmdefault}{\mddefault}{\updefault}$y_2$}}}}
\put(2953,223){\makebox(0,0)[lb]{\smash{{\SetFigFont{8}{9.6}{\rmdefault}{\mddefault}{\updefault}$y_3$}}}}
\put(628,748){\makebox(0,0)[lb]{\smash{{\SetFigFont{8}{9.6}{\rmdefault}{\mddefault}{\updefault}$1-y_1$}}}}
\put(1828,748){\makebox(0,0)[lb]{\smash{{\SetFigFont{8}{9.6}{\rmdefault}{\mddefault}{\updefault}$1-y_2$}}}}
\put(3028,748){\makebox(0,0)[lb]{\smash{{\SetFigFont{8}{9.6}{\rmdefault}{\mddefault}{\updefault}$1-y_3$}}}}
\put(628,2023){\makebox(0,0)[lb]{\smash{{\SetFigFont{8}{9.6}{\rmdefault}{\mddefault}{\updefault}$x_1$}}}}
\put(1828,2023){\makebox(0,0)[lb]{\smash{{\SetFigFont{8}{9.6}{\rmdefault}{\mddefault}{\updefault}$x_2$}}}}
\put(3028,2023){\makebox(0,0)[lb]{\smash{{\SetFigFont{8}{9.6}{\rmdefault}{\mddefault}{\updefault}$x_3$}}}}
\end{picture}
}
   \caption{An infinite Markov chain on which
     $\simclo \ \not\subseteq \ \approx^\pos$, where $x_i$'s and
     $y_i$'s represent the probabilities that the corresponding edge
     is taken.}
  \label{fig2}
\end{figure}
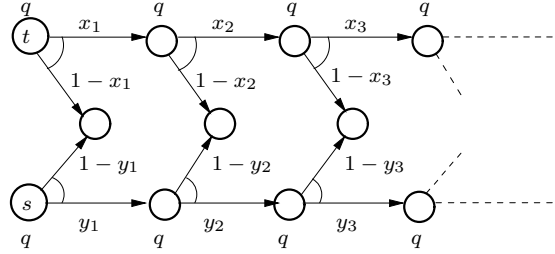

\begin{proof}
{\em Assertion 1.} 
We prove that on a finite, alternating MDP, the counterpositive holds: 
if $s \not\approx^\pos t$, then $s \not\simclo t$. 
Let $s$ and $t$ be two states such that $s \not\approx^\pos t$. 
Then, there must be a formula $\varphi$ in $\qrctl^\pos$ that
distinguishes $s$ from $t$. 
From this formula, we derive a formula $f(\varphi)$ in $\atl$ that
distinguishes $s$ from $t$.

We proceed by structural induction on $\varphi$, starting from the
inner part of the formula and replacing successive parts that are in
the scope of a path quantifier by their $\atl$ version.
The cases where $\varphi$ is an atomic proposition, or a boolean
combination of formulas are trivial. 
Using (\ref{eq-dual}), we reduce $\qrctl^\pos$-formulas
that involve a $\forall$ operator to formulas that only involve
the $\exists$ operator.
Lemma~\ref{lem-trans1} provides
translations for all such formulas, except those of type
$\exists^\almost (\phi \until \psi)$. 
For instance, (\ref{eq-transl-pos-until}) leads to 
$f(\exists^\pos  \phi \until \psi) = \plbr{1,p} f(\phi) \until f(\psi)$.
In order to translate a formula of the form $\gamma = \exists^\almost
(\phi \until \psi)$, we translate the evaluation of the nested
$\mu$-calculus formula (\ref{sem-until-almost}) into the evaluation of
a nested \atl\ formula, as follows. 
Define the set of formulas 
$\set{\alpha_{i,j} \mid 0 \leq i, j \leq n}$, where $n = |S|$ is the
number of states of the AMDP, via the following clauses: 
\begin{align*}
  \forall i \in [0..n]: \quad & \alpha_{i,0} = \false \\
  \forall j \in [1..n]: \quad & \alpha_{0,j} = \true \\
  \begin{split} 
    \forall i \in [1..n] \qdot \forall j \in [0..n-1]: \quad 
  & \\ \alpha_{i,j+1} = f(\psi) \oder \bigl(f(\phi) \und 
	&F_\apre(\alpha_{i-1,n}, \alpha_{i,j}) \bigr).
  \end{split}
\end{align*}
From Lemma~\ref{lem-trans3}, the above set of formulas encodes the
iterative evaluation of the nested fixpoint (\ref{sem-until-almost}),
so that we have $\sem{\alpha_{n,n}} = \sem{\gamma}$, and we can define
$f(\gamma) = \alpha_{n,n}$. 
This concludes the translation.

\medskip 

\noindent {\em Assertion 2.} 
Consider the MDP shown in Figure~\ref{fig4}.
The states $s$
and $t$ are such that $(s,t) \in \,\simclo$.
However, $s \models \exists^\almost (\Diamond q)$ (consider the
strategy that plays always $a$), whereas $t \not\models \exists^\almost(\Diamond q)$.

\medskip 

\noindent {\em Assertion 3.} 
Consider the infinite AMDP shown in Figure~\ref{fig2}.
All states are probabilistic states, i.e. $S_1 = \emptyset$.
For all $i>0$, we set $x_i = \frac{1}{2}$
and $y_i = 2^{-\frac{1}{2^i}}$, 
so that $\prod_{i>0} \, x_i = 0$ and $\prod_{i>0} \, y_i = \frac{1}{2}$.
It is easy to see that $s \simclo t$. 
However, 
$s \models \exists^\pos (\Box q)$ and $t \not\models \exists^\pos(
\Box q)$.
\end{proof}

The example in Figure~\ref{fig4} also shows that on non-alternating MDPs,
unlike on alternating ones (see Lemma~\ref{lem-trans3}), 
the $\apre$ operator cannot be encoded in $\atl$.
If we were able to encode $\apre$ in $\atl$, by proceeding as in the
proof of the first assertion, given two states $s$, $t$ 
with $s \not\approx^\pos t$, we could construct an $\atl$ formula
distinguishing $s$ from $t$.

As a corollary to Theorems \ref{lemm:possub} and \ref{lemm:simsub}, we
have that on finite, alternating MDPs, the
equivalences induced by $\atl$ and $\qrctl$ coincide.
Thus the discrete graph theoretic algorithms to compute equivalences 
for $\atl$ can be used to compute the $\qrctl$ equivalences for finite AMDPs.

\begin{cor}\label{thrm:pos}
For finite AMDPs, we have $\approx^\pos \ =\ \simclo$.
\end{cor}

\section{Computing $\qrctl$ Equivalences}
\label{res2}

In this section, we take advantage of the results obtained in Section
\ref{res1} to derive algorithms to compute $\approx^\pos$ and
$\approx^\sure$ for AMDPs. We also provide an algorithm to compute
those relations on non-alternating MDPs.

\subsection{Alternating MDPs} 

Corollary \ref{thrm:pos} immediately provides an algorithm for the
computation of the $\qrctl$ equivalences on AMDPs, via the computation
of the $\atl$ equivalences (interpreting nondeterminism and
probability as the two players).
In particular, the partition-refinement algorithms presented in
\cite{ATL02} can be directly applied to the problem. 
This yields the following result. 

\begin{thm} 
The two problems of computing $\approx^\pos$ and $\approx^\sure$
on finite AMDPs are PTIME-complete. 
\end{thm}

\begin{proof}
Consider a turn-based game and consider the AMDP obtained from the 
game assigning uniform transition probabilities to all out-going edges
from a player~2 state.
Then the $2$-player game interpretation of the AMDP coincides with 
the original turn-based game.
The result then follows from Corollary~\ref{thrm:pos}, and from the
PTIME-completeness of ATL model checking and 
computing $\simclo$~\cite{ATL02}. 
\end{proof}

\subsection{Non-Alternating MDPs} 

For the general case of {\em non-alternating\/} MDPs, on the other hand,
the situation is not nearly as simple. 
First, let us dispel the belief that, in order to compute
$\approx^\pos$ on a non-alternating MDP, we can convert the MDP into an alternating one, 
compute $\approx^\pos$ via $\simclo$ (using
Corollary~\ref{thrm:pos}) on the alternating one, and then somehow
obtain $\approx^\pos$ on the original non-alternating MDP. 
The following example shows that this, in general, is not possible. 

\begin{figure}[t]
\centering 
\mbox{\hspace*{-1em}
\subfigure[A non-alternating MDP.]{
\label{fig-conc-1}
\setlength{\unitlength}{0.00050000in}
\begingroup\makeatletter\ifx\SetFigFont\undefined%
\gdef\SetFigFont#1#2#3#4#5{%
  \reset@font\fontsize{#1}{#2pt}%
  \fontfamily{#3}\fontseries{#4}\fontshape{#5}%
  \selectfont}%
\fi\endgroup%
{\renewcommand{\dashlinestretch}{30}
\begin{picture}(2521,2129)(0,-10)
\put(2536.000,-218.000){\arc{2757.716}{3.5322}{4.3218}}
\blacken\path(1914.712,979.356)(2011.000,1057.000)(1889.502,1033.802)(1914.712,979.356)
\put(-14.000,-218.000){\arc{2757.716}{5.1030}{5.8926}}
\blacken\path(632.498,1033.802)(511.000,1057.000)(607.288,979.356)(632.498,1033.802)
\put(1254.500,429.351){\arc{205.709}{3.7133}{5.7115}}
\put(2336.000,1254.000){\arc{353.553}{2.9997}{7.9959}}
\blacken\path(2156.248,1352.602)(2161.000,1229.000)(2214.973,1340.297)(2156.248,1352.602)
\put(186.000,1232.000){\arc{353.553}{1.4289}{6.4251}}
\blacken\path(307.027,1318.297)(361.000,1207.000)(365.752,1330.602)(307.027,1318.297)
\put(2161,1079){\ellipse{300}{300}}
\put(1261,1957){\ellipse{300}{300}}
\put(1261,157){\ellipse{300}{300}}
\put(361,1057){\ellipse{300}{300}}
\path(1411,1957)(2161,1207)
\blacken\path(2054.934,1270.640)(2161.000,1207.000)(2097.360,1313.066)(2054.934,1270.640)
\path(1111,1957)(361,1207)
\blacken\path(424.640,1313.066)(361.000,1207.000)(467.066,1270.640)(424.640,1313.066)
\path(1111,157)(361,907)
\blacken\path(467.066,843.360)(361.000,907.000)(424.640,800.934)(467.066,843.360)
\path(1418,182)(2168,932)
\blacken\path(2104.360,825.934)(2168.000,932.000)(2061.934,868.360)(2104.360,825.934)
\put(511,307){\makebox(0,0)[lb]{\smash{{\SetFigFont{7}{8.4}{\rmdefault}{\mddefault}{\updefault}$a$}}}}
\put(1861,307){\makebox(0,0)[lb]{\smash{{\SetFigFont{7}{8.4}{\rmdefault}{\mddefault}{\updefault}$b$}}}}
\put(1186,607){\makebox(0,0)[lb]{\smash{{\SetFigFont{7}{8.4}{\rmdefault}{\mddefault}{\updefault}$c$}}}}
\put(1861,1657){\makebox(0,0)[lb]{\smash{{\SetFigFont{7}{8.4}{\rmdefault}{\mddefault}{\updefault}$b$}}}}
\put(511,1657){\makebox(0,0)[lb]{\smash{{\SetFigFont{7}{8.4}{\rmdefault}{\mddefault}{\updefault}$a$}}}}
\put(104,728){\makebox(0,0)[lb]{\smash{{\SetFigFont{7}{8.4}{\rmdefault}{\mddefault}{\updefault}$r$}}}}
\put(2213,757){\makebox(0,0)[lb]{\smash{{\SetFigFont{7}{8.4}{\rmdefault}{\mddefault}{\updefault}$q$}}}}
\put(286,987){\makebox(0,0)[lb]{\smash{{\SetFigFont{7}{8.4}{\rmdefault}{\mddefault}{\updefault}$u$}}}}
\put(1186,1900){\makebox(0,0)[lb]{\smash{{\SetFigFont{7}{8.4}{\rmdefault}{\mddefault}{\updefault}$s$}}}}
\put(1206,93){\makebox(0,0)[lb]{\smash{{\SetFigFont{7}{8.4}{\rmdefault}{\mddefault}{\updefault}$s'$}}}}
\put(2106,1022){\makebox(0,0)[lb]{\smash{{\SetFigFont{7}{8.4}{\rmdefault}{\mddefault}{\updefault}$t$}}}}
\end{picture}
}
}
\subfigure[An alternating MDP.]{
\label{fig-conc-2}
\setlength{\unitlength}{0.00050000in}
\begingroup\makeatletter\ifx\SetFigFont\undefined%
\gdef\SetFigFont#1#2#3#4#5{%
  \reset@font\fontsize{#1}{#2pt}%
  \fontfamily{#3}\fontseries{#4}\fontshape{#5}%
  \selectfont}%
\fi\endgroup%
{\renewcommand{\dashlinestretch}{30}
\begin{picture}(3318,2281)(0,-10)
\put(168.375,1133.000){\arc{318.750}{0.4900}{5.7932}}
\blacken\path(235.480,958.527)(309.000,1058.000)(197.319,1004.828)(235.480,958.527)
\put(3149.625,1133.000){\arc{318.750}{3.6315}{8.9348}}
\blacken\path(3120.681,1004.828)(3009.000,1058.000)(3082.520,958.527)(3120.681,1004.828)
\put(1220.250,-37.000){\arc{2640.096}{4.2310}{5.0512}}
\blacken\path(705.266,1210.672)(609.000,1133.000)(730.492,1156.232)(705.266,1210.672)
\put(2097.750,-37.000){\arc{2640.096}{4.3736}{5.1938}}
\blacken\path(2587.508,1156.232)(2709.000,1133.000)(2612.734,1210.672)(2587.508,1156.232)
\put(1659.000,1170.500){\arc{375.000}{3.7851}{5.6397}}
\put(459,2033){\ellipse{750}{450}}
\put(2859,2033){\ellipse{750}{450}}
\put(2859,233){\ellipse{750}{450}}
\put(1659,983){\ellipse{750}{450}}
\put(1659,2033){\ellipse{300}{300}}
\put(1659,233){\ellipse{300}{300}}
\put(459,233){\ellipse{750}{450}}
\put(459,1133){\ellipse{300}{300}}
\put(2859,1133){\ellipse{300}{300}}
\path(1659,383)(1659,758)
\blacken\path(1689.000,638.000)(1659.000,758.000)(1629.000,638.000)(1689.000,638.000)
\path(1509,233)(834,233)
\blacken\path(954.000,263.000)(834.000,233.000)(954.000,203.000)(954.000,263.000)
\path(459,458)(459,983)
\blacken\path(489.000,863.000)(459.000,983.000)(429.000,863.000)(489.000,863.000)
\path(1809,2033)(2484,2033)
\blacken\path(2364.000,2003.000)(2484.000,2033.000)(2364.000,2063.000)(2364.000,2003.000)
\path(1509,2033)(834,2033)
\blacken\path(954.000,2063.000)(834.000,2033.000)(954.000,2003.000)(954.000,2063.000)
\path(459,1808)(459,1283)
\blacken\path(429.000,1403.000)(459.000,1283.000)(489.000,1403.000)(429.000,1403.000)
\path(2859,1808)(2859,1283)
\blacken\path(2829.000,1403.000)(2859.000,1283.000)(2889.000,1403.000)(2829.000,1403.000)
\path(1809,233)(2484,233)
\blacken\path(2364.000,203.000)(2484.000,233.000)(2364.000,263.000)(2364.000,203.000)
\path(2859,458)(2859,983)
\blacken\path(2889.000,863.000)(2859.000,983.000)(2829.000,863.000)(2889.000,863.000)
\put(1359,908){\makebox(0,0)[lb]{\smash{{\SetFigFont{7}{8.4}{\rmdefault}{\mddefault}{\updefault}$\tuple{s',c}$}}}}
\put(534,833){\makebox(0,0)[lb]{\smash{{\SetFigFont{7}{8.4}{\rmdefault}{\mddefault}{\updefault}$r$}}}}
\put(2559,833){\makebox(0,0)[lb]{\smash{{\SetFigFont{7}{8.4}{\rmdefault}{\mddefault}{\updefault}$q$}}}}
\put(1584,163){\makebox(0,0)[lb]{\smash{{\SetFigFont{7}{8.4}{\rmdefault}{\mddefault}{\updefault}$s'$}}}}
\put(1584,1990){\makebox(0,0)[lb]{\smash{{\SetFigFont{7}{8.4}{\rmdefault}{\mddefault}{\updefault}$s$}}}}
\put(2586,1985){\makebox(0,0)[lb]{\smash{{\SetFigFont{7}{8.4}{\rmdefault}{\mddefault}{\updefault}$\tuple{s,b}$}}}}
\put(2586,190){\makebox(0,0)[lb]{\smash{{\SetFigFont{7}{8.4}{\rmdefault}{\mddefault}{\updefault}$\tuple{s',b}$}}}}
\put(192,185){\makebox(0,0)[lb]{\smash{{\SetFigFont{7}{8.4}{\rmdefault}{\mddefault}{\updefault}$\tuple{s',a}$}}}}
\put(205,1984){\makebox(0,0)[lb]{\smash{{\SetFigFont{7}{8.4}{\rmdefault}{\mddefault}{\updefault}$\tuple{s,a}$}}}}
\put(397,1090){\makebox(0,0)[lb]{\smash{{\SetFigFont{7}{8.4}{\rmdefault}{\mddefault}{\updefault}$u$}}}}
\put(2804,1083){\makebox(0,0)[lb]{\smash{{\SetFigFont{7}{8.4}{\rmdefault}{\mddefault}{\updefault}$t$}}}}
\end{picture}
}
}
}
\caption{MDPs illustrating how separating nondeterministic and
  probabilistic choice does not help to compute $\approx^\pos$.}
\end{figure}
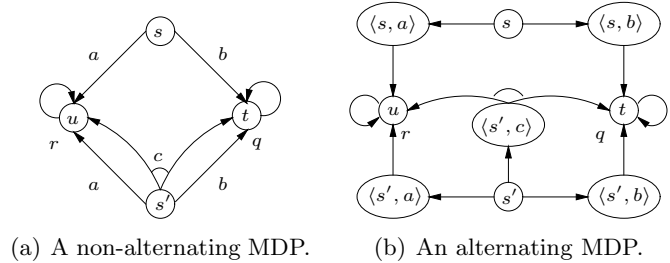

\begin{exa}{} 
Consider the MDP depicted in Figure~\ref{fig-conc-1},
where the set of predicates is $\AP=\set{q,r}$.
We have $s \approx^\pos s'$.  Indeed, the only difference between 
$s$ and $s'$ is that at state $s'$ the action $c$ is available:
since $c$ is a convex combination of $a$ and $b$, 
$s$ and $s'$ are probabilistically bisimilar in the sense of
\cite{SL94}, and thus also related by $\approx^\pos$. 
We transform this MDP into an alternating one by adding, for each state
$s$ and each $a \in \mov(s)$, a state $\tuple{s,a}$ which represents the
decision of choosing $a$ at $s$; the result is depicted in 
Figure~\ref{fig-conc-2}. 
In this AMDP, however, the state $\tuple{s',c}$ has no equivalent, as
it satisfies both $\exists^\pos \nextop q$ and $\exists^\pos \nextop r$. 
Therefore, on this AMDP we have $s \ \not\approx^\pos s'$, as
witnessed by the formula
$\exists^\sure \nextop ((\exists^\pos \nextop q) \und (\exists^\pos \nextop r))$. 
\end{exa}

\noindent
As the example illustrates, the problem is that once nondeterminism
and probability are separated into different states, the
distinguishing power of $\approx^\pos$ increases, so that computing
$\simclo$ on the resulting alternating MDP does not help to compute
$\approx^\pos$ on the original non-alternating one. 

\subsubsection*{Failure of local partition refinement} 

Simulation and bisimulation relations can be computed via 
partition refinement algorithms that consider, at each step, the 
{\em 1-neighbourhood\/} of each state: that is, the set of states
reachable from a given state in one step \cite{Milner90}. 
We call such algorithms {\em 1-neighbourhood partition refinements.}
Here, we show a general result: no 1-neighbourhood partition
refinement algorithm exists for $\approx^\pos$ on non-alternating MDPs. 

We make this notion precise as follows. 
Consider an MDP $G=(S,\Acts,\mov,\trans, \lab{\cdot})$, 
together with an equivalence relation $\sim$ on $S$. 
Intuitively, two states are 1-neighbourhood isomorphic up to $\sim$
if their 1-step future looks identical, up to the equivalence
$\sim$. 
Formally, we say that two states $s, t \in S$ are 
{\em 1-neighbourhood isomorphic up to $\sim$,} written 
$s \stackrel{1}{\sim} t$, iff $s \sim t$,
and if there is a bijection $R$ between 
$E(s)$ and $E(t)$, and a bijection $\hat{R}$
between $\mov(s)$ and $\mov(t)$, which preserve $\sim$ and the
transition probabilities.  
Precisely, we require that: 
\begin{enumerate}[$\bullet$] 
\item if $s' \in E(s)$ and $t' \in E(t)$ with $s' \, R \, t'$, then
  $s' \sim t'$; 
\item if $a \in \mov(s)$ and $b \in \mov(t)$ with $a \, \hat{R} \, b$, 
  then for all $s' \in E(s)$ and $t' \in E(t)$ with $s' \, R \, t'$, 
  we have $\trans(s,a)(s') = \trans(t,b)(t')$. 
\end{enumerate}
Let $\partitions{S}$ be the set of equivalence relations on $S$. 
A {\em partition refinement operator\/} 
$f: \partitions{S} \mapsto \partitions{S}$ is an operator such that,
for all $\sim \ \in \partitions{S}$, we have 
$f(\sim)$ is finer than $\sim$. 
We say that a partition operator {\em computes\/} a relation $\approx$
if we have $\approx \ = \lim_{n \go \infty} f^n(\sim_{\textit{pred}})$,
where $f^n$ denotes $n$ repeated applications of $f$
and $s \sim_{{\text{\textit{pred}}}} t$ iff $\lab{s} = \lab{t}$. 

We say that a partition refinement operator $f$ is 
{\em 1-neighbourhood\/} if it refines an equivalence relation $\sim$
on the basis of the 1-neighbourhood of the states, treating in the
same fashion states whose 1-neighbourhoods are isomorphic up to
$\sim$.  
Precisely, $f$ is {\em 1-neighbourhood\/} if, 
for all $\sim\,\in \partitions{S}$ and 
for all $s, s', t, t' \in S$ with  
$s \stackrel{1}{\sim} s'$, 
$t \stackrel{1}{\sim} t'$, 
we have 
either $(s,t), (s', t') \in f(\sim)$, 
or $(s,t), (s', t') \not\in f(\sim)$.
We can now state the non-existence of 1-neighbourhood refinement
operators for $\approx^\pos$ as follows. 


\begin{figure}[h]
  \centering
  \setlength{\unitlength}{0.00037500in}
\begingroup\makeatletter\ifx\SetFigFont\undefined%
\gdef\SetFigFont#1#2#3#4#5{%
  \reset@font\fontsize{#1}{#2pt}%
  \fontfamily{#3}\fontseries{#4}\fontshape{#5}%
  \selectfont}%
\fi\endgroup%
{\renewcommand{\dashlinestretch}{30}
\begin{picture}(8494,3258)(0,-10)
\put(532.000,987.000){\arc{750.000}{5.3559}{6.2832}}
\put(1169.500,649.500){\arc{530.330}{1.4289}{3.2835}}
\path(532,1587)(757,1287)
\path(907,987)(907,687)
\blacken\path(869.500,837.000)(907.000,687.000)(944.500,837.000)(869.500,837.000)
\put(3907.000,2437.000){\arc{500.000}{5.3559}{10.3521}}
\blacken\path(3700.687,2493.003)(3757.000,2637.000)(3639.549,2536.444)(3700.687,2493.003)
\put(3907.000,737.000){\arc{500.000}{2.2143}{7.2105}}
\blacken\path(4113.313,680.997)(4057.000,537.000)(4174.451,637.556)(4113.313,680.997)
\put(3907,2787){\ellipse{450}{450}}
\put(3907,387){\ellipse{450}{450}}
\path(2332,1587)(2557,1287)
\put(1882.000,987.000){\arc{750.000}{3.1416}{4.0689}}
\put(1769.500,649.500){\arc{530.330}{1.4289}{3.2835}}
\path(1882,1587)(1657,1287)
\path(1507,987)(1507,687)
\blacken\path(1469.500,837.000)(1507.000,687.000)(1544.500,837.000)(1469.500,837.000)
\put(307.000,1978.071){\arc{567.856}{2.1274}{7.2974}}
\blacken\path(27.829,1821.977)(157.000,1737.000)(83.015,1872.766)(27.829,1821.977)
\put(588.250,1587.000){\arc{487.500}{5.1072}{7.4592}}
\put(8519.500,1399.500){\arc{2855.915}{4.1900}{4.4738}}
\put(2107.000,1978.071){\arc{567.856}{2.1274}{7.2974}}
\blacken\path(1827.829,1821.977)(1957.000,1737.000)(1883.015,1872.766)(1827.829,1821.977)
\put(2332.000,987.000){\arc{750.000}{5.3559}{6.2832}}
\put(2332.000,2187.000){\arc{750.000}{6.2832}{7.2105}}
\put(2969.500,649.500){\arc{530.330}{1.4289}{3.2835}}
\put(2969.500,2524.500){\arc{530.330}{2.9997}{4.8543}}
\put(2388.250,1587.000){\arc{487.500}{5.1072}{7.4592}}
\put(2332.000,987.000){\arc{750.000}{5.3559}{6.2832}}
\put(2332.000,2187.000){\arc{750.000}{6.2832}{7.2105}}
\put(2969.500,649.500){\arc{530.330}{1.4289}{3.2835}}
\put(2969.500,2524.500){\arc{530.330}{2.9997}{4.8543}}
\put(2388.250,1587.000){\arc{487.500}{5.1072}{7.4592}}
\put(5707.000,1978.071){\arc{567.856}{2.1274}{7.2974}}
\blacken\path(5930.985,1872.766)(5857.000,1737.000)(5986.171,1821.977)(5930.985,1872.766)
\put(5482.000,987.000){\arc{750.000}{3.1416}{4.0689}}
\put(5482.000,2187.000){\arc{750.000}{2.2143}{3.1416}}
\put(4844.500,649.500){\arc{530.330}{6.1413}{7.9959}}
\put(4844.500,2524.500){\arc{530.330}{4.5705}{6.4251}}
\put(5482.000,987.000){\arc{750.000}{3.1416}{4.0689}}
\put(5482.000,2187.000){\arc{750.000}{2.2143}{3.1416}}
\put(4844.500,649.500){\arc{530.330}{6.1413}{7.9959}}
\put(4844.500,2524.500){\arc{530.330}{4.5705}{6.4251}}
\put(7507.000,1978.071){\arc{567.856}{2.1274}{7.2974}}
\blacken\path(7730.985,1872.766)(7657.000,1737.000)(7786.171,1821.977)(7730.985,1872.766)
\put(7282.000,987.000){\arc{750.000}{3.1416}{4.0689}}
\put(6644.500,649.500){\arc{530.330}{6.1413}{7.9959}}
\put(7225.750,1587.000){\arc{487.500}{1.9656}{4.3176}}
\put(7282.000,987.000){\arc{750.000}{3.1416}{4.0689}}
\put(6644.500,649.500){\arc{530.330}{6.1413}{7.9959}}
\put(7225.750,1587.000){\arc{487.500}{1.9656}{4.3176}}
\put(1657.000,1362.000){\arc{335.410}{2.0344}{3.6052}}
\put(8219.500,2524.500){\arc{530.330}{4.5705}{6.4251}}
\put(7308.670,1254.070){\arc{451.592}{5.7848}{8.3360}}
\blacken\path(7566.776,1219.406)(7507.000,1362.000)(7492.641,1208.052)(7566.776,1219.406)
\put(8254.077,1653.397){\arc{456.170}{6.2455}{8.1755}}
\put(7884.577,1416.039){\arc{503.450}{6.1479}{7.2283}}
\put(7732.000,987.000){\arc{750.000}{5.3559}{6.2832}}
\put(7786.190,707.810){\arc{642.968}{0.0648}{1.5060}}
\put(6601.401,2481.401){\arc{611.301}{4.7307}{6.2649}}
\put(7293.159,2198.159){\arc{781.258}{2.2200}{3.0925}}
\put(6044.500,649.500){\arc{530.330}{6.1413}{7.9959}}
\put(5816.098,936.218){\arc{978.289}{5.4835}{6.1874}}
\put(6155.955,1362.000){\arc{337.281}{5.8223}{7.4060}}
\put(5427.192,1586.039){\arc{490.387}{2.0369}{4.3137}}
\put(5424.921,1587.553){\arc{485.843}{2.0589}{4.3199}}
\put(510.884,1265.884){\arc{450.808}{1.1303}{3.5821}}
\blacken\path(329.518,1209.032)(307.000,1362.000)(254.884,1216.432)(329.518,1209.032)
\put(1204.063,2489.937){\arc{594.155}{3.1317}{4.7223}}
\put(343.926,2281.037){\arc{1141.745}{0.1655}{0.7618}}
\put(5707,1587){\ellipse{450}{450}}
\put(307,1587){\ellipse{450}{450}}
\put(2107,1587){\ellipse{450}{450}}
\put(7507,1587){\ellipse{450}{450}}
\path(1207,2787)(3682,2787)
\path(1207,387)(3682,387)
\path(532,1587)(757,1887)
\path(907,2187)(907,2487)
\blacken\path(944.500,2337.000)(907.000,2487.000)(869.500,2337.000)(944.500,2337.000)
\path(7732,1587)(8182,1437)
\blacken\path(6047.782,1764.474)(5932.000,1662.000)(6082.383,1697.932)(6047.782,1764.474)
\path(5932,1662)(7807,2637)
\path(4132,2787)(6607,2787)
\path(4132,387)(7807,387)
\path(1882,1587)(607,1062)
\path(2332,1587)(2557,1887)
\path(2332,1587)(2557,1887)
\path(2332,1587)(2557,1287)
\path(2707,2187)(2707,2487)
\blacken\path(2744.500,2337.000)(2707.000,2487.000)(2669.500,2337.000)(2744.500,2337.000)
\path(2707,987)(2707,687)
\blacken\path(2669.500,837.000)(2707.000,687.000)(2744.500,837.000)(2669.500,837.000)
\path(2332,1587)(2557,1887)
\path(2332,1587)(2557,1287)
\path(2707,2187)(2707,2487)
\blacken\path(2744.500,2337.000)(2707.000,2487.000)(2669.500,2337.000)(2744.500,2337.000)
\path(2707,987)(2707,687)
\blacken\path(2669.500,837.000)(2707.000,687.000)(2744.500,837.000)(2669.500,837.000)
\path(5482,1587)(5257,1887)
\path(5482,1587)(5257,1887)
\path(5482,1587)(5257,1287)
\path(5107,2187)(5107,2487)
\blacken\path(5144.500,2337.000)(5107.000,2487.000)(5069.500,2337.000)(5144.500,2337.000)
\path(5107,987)(5107,687)
\blacken\path(5069.500,837.000)(5107.000,687.000)(5144.500,837.000)(5069.500,837.000)
\path(5482,1587)(5257,1887)
\path(5482,1587)(5257,1287)
\path(5107,2187)(5107,2487)
\blacken\path(5144.500,2337.000)(5107.000,2487.000)(5069.500,2337.000)(5144.500,2337.000)
\path(5107,987)(5107,687)
\blacken\path(5069.500,837.000)(5107.000,687.000)(5144.500,837.000)(5069.500,837.000)
\path(7282,1587)(7057,1887)
\path(7282,1587)(7057,1887)
\path(7282,1587)(7057,1287)
\path(6907,987)(6907,687)
\blacken\path(6869.500,837.000)(6907.000,687.000)(6944.500,837.000)(6869.500,837.000)
\path(7282,1587)(7057,1887)
\path(7282,1587)(7057,1287)
\path(6906,2182)(6907,2487)
\blacken\path(6944.008,2336.878)(6907.000,2487.000)(6869.008,2337.124)(6944.008,2336.878)
\path(6907,987)(6907,687)
\blacken\path(6869.500,837.000)(6907.000,687.000)(6944.500,837.000)(6869.500,837.000)
\path(7204,1054)(5929,1579)
\path(8482,2487)(8482,1662)
\path(7732,1587)(7957,1287)
\path(8107,987)(8107,687)
\blacken\path(8069.500,837.000)(8107.000,687.000)(8144.500,837.000)(8069.500,837.000)
\path(7282,1587)(7057,1287)
\path(5482,1587)(5257,1287)
\path(5932,1587)(6157,1287)
\path(6303,988)(6307,687)
\blacken\path(6267.510,836.488)(6307.000,687.000)(6342.504,837.485)(6267.510,836.488)
\put(4057,3012){\makebox(0,0)[lb]{\smash{{\SetFigFont{8}{9.6}{\rmdefault}{\mddefault}{\updefault}$q$}}}}
\put(4132,87){\makebox(0,0)[lb]{\smash{{\SetFigFont{8}{9.6}{\rmdefault}{\mddefault}{\updefault}$r$}}}}
\put(7405,1525){\makebox(0,0)[lb]{\smash{{\SetFigFont{8}{9.6}{\rmdefault}{\mddefault}{\updefault}$s_4$}}}}
\put(5572,1512){\makebox(0,0)[lb]{\smash{{\SetFigFont{8}{9.6}{\rmdefault}{\mddefault}{\updefault}$s_3$}}}}
\put(1992,1519){\makebox(0,0)[lb]{\smash{{\SetFigFont{8}{9.6}{\rmdefault}{\mddefault}{\updefault}$s_2$}}}}
\put(192,1525){\makebox(0,0)[lb]{\smash{{\SetFigFont{8}{9.6}{\rmdefault}{\mddefault}{\updefault}$s_1$}}}}
\end{picture}
}
  \caption{ 
   MDP showing the lack of 1-neighbourhood refinement
   operators.}
  \label{fig-1neighbour}
\end{figure}
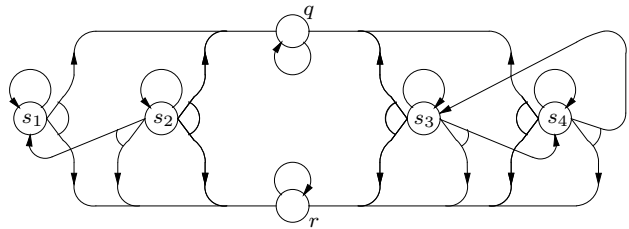

\begin{thm} 
\label{theo-1-neigh}
There is no 1-neighbourhood partition refinement operator which
computes $\approx^\pos$ on all MDPs. 
\end{thm}

\begin{proof}
Consider the states $s_1, s_2, s_3, s_4$ of the MDP 
depicted in Figure~\ref{fig-1neighbour}, 
and take $\sim \ = \ \sim_{{\text{\textit{pred}}}}$.
Let $f$ be any 1-neighbourhood partition refinement operator.
From $s_1 \sim s_2 \sim s_3 \sim s_4$, we can see that 
$s_2 \stackrel{1}{\sim} s_3 \stackrel{1}{\sim} s_4$. 
Let $\sim' \ = f(\sim)$.
Considering the pairs $(s_1,s_2)$, $(s_1,s_3)$, and $(s_1,s_4)$
in the definition of 1-neighbourhood partition refinement operator, we
have that $\sim'$ satisfies one of the following two cases: 
\begin{enumerate}[(1)] 
\item $s_1 \not\sim' s_2$ and $s_1 \not\sim' s_3$ and 
  $s_1 \not\sim' s_4$, 
\item $s_1 \sim' s_2$ and $s_1 \sim' s_3$ and  $s_1 \sim' s_4$. 
\end{enumerate}
In the first case, the partition refinement terminates with a
relation $\sim''$ such that $s_1 \not\sim'' s_2$. 
This is incorrect, since we can prove by induction on the length of $\qrctl^\pos$
formulas that no such formula distinguishes $s_1$ from $s_2$,
so that $s_1 \approx^\pos s_2$. 
In the second case, the partition refinement terminates with a
relation $\sim''$ such that $s_1 \sim'' s_3$. 
This is also incorrect, since the formula $\exists^\almost \diam r$ is a
witness to $s_1 \not\approx^\pos s_3$. 
We conclude that a 1-neighbourhood partition refinement operator
cannot compute $\approx^\pos$. 
\end{proof}

To give an algorithm for the computation of $\approx^\pos$, 
given two sets of states $C_1$ and $C_2$, let:
\begin{align*}
  \begin{split}
  U(C_1, C_2) & = \set{ \infpat=\seq{s_0,s_1,\ldots} \mid 
    \exists j\geq 0 \qdot s_j \in C_2 \text{ and } 
\forall\, 0 \leq i < j \qdot s_i \in C_1} 
  \end{split} \\
  EU^\almost(C_1, C_2) & = \set{s \in S \mid 
    \exists \straa \in \Straa.\ \Prb_s^{\straa}( U(C_1,C_2) ) = 1}.
\end{align*}
Intuitively, if $C_1 = \sem{\phi_1}$ and $C_2 = \sem{\phi_2}$
for two \qrctl\ formulas $\phi_1$ and $\phi_2$, then
$EU^\almost(C_1, C_2)$ is $\sem{\exists^\almost (\phi_1 \,\until\, \phi_2)}$.

\noindent
We say that an equivalence relation $\simeq$ is {\em $1,p,EU$-stable\/}
if, for all unions $C_1, C_2$ of equivalence classes with respect to
$\simeq$, and for all $s, t \in S$ with $s \simeq t$, we have: 
\begin{enumerate}[(1)]
\item $s \in \pre(C_1)$ iff $t \in \pre(C_1)$; 
\item $s \in \cpre(C_1)$ iff $t \in \cpre(C_1)$; 
\item $s \in EU^\almost (C_1, C_2)$ iff $t \in EU^\almost (C_1, C_2)$.
\end{enumerate}
Let $\simuntclo$ be the coarsest equivalence relation that is
$1,p,EU$-stable. 
We show that $\simuntclo$ coincides with $\approx^\pos$. 

\begin{thm}{}
\label{theo-1-neigh2}
For all finite MDPs, we have $\simuntclo \ = \ \approx^\pos$.
\end{thm}

\begin{proof}
We prove containment in the two directions. 

\medskip 

\noindent
$\simuntclo \ \subseteq \ \approx^\pos$.
This statement is equivalent to saying that
for all formulas $\phi$ in $\qrctl^\pos$, $\sem{\phi}$ is
the union of classes in $S/\!\simuntclo$.
Let $s$ and $t$ be two states such that $s \not\approx^\pos t$,
and let $\phi$ be a formula from $\qrctl^\pos$ such that $s \sat \phi$
and $t \not\sat \phi$.
We show by structural induction on $\phi$ that $s \not\simuntclo t$. 
The cases where $\phi$ is a proposition, or the boolean combination
of formulas are trivial.
All other cases follow as in the proof of the first part of 
Theorem~\ref{lemm:simsub}, except for the case 
$\phi = \exists^\almost (\phi_1 \until \phi_2)$. 
For $\phi = \exists^\almost (\phi_1 \until \phi_2)$, 
we have $s \in EU^\almost(\sem{\phi_1},\sem{\phi_2})$,
while $t \not\in EU^\almost(\sem{\phi_1},\sem{\phi_2})$.
By inductive hypothesis, we can assume that $\sem{\phi_1}$ and $\sem{\phi_2}$
are unions of classes in $S/\!\simuntclo$.
So, $(s,t) \not\in\, \simuntclo$.  

\medskip 

\noindent 
$\approx^\pos \ \subseteq \ \simuntclo$. 
The proof follows the same idea of the proof of the first part of
Theorem~\ref{lemm:possub}.  The only modification needed is in the
inductive definition of the set of formulas: we take here 
$
  \Psi_{k+1} = \clos(\Psi_k \union 
  \set{\exists^\pos \nextop \psi, 
    \exists^\sure \nextop \psi, 
    \exists^\almost \psi \until \psi' 
    \mid \psi, \psi' \in \Psi_k})
$.
\end{proof}

The following theorem provides an upper bound for the complexity of
computing $\approx^\pos$ on MDPs. 
The PTIME-completeness of ordinary simulation \cite{bisim-hard91} provides a
lower bound, but no tight lower bound is known. 

\begin{thm} \label{theo-conp}
The problem of deciding whether $s\approx^\pos t$ for two states $s$ and $t$
of an MDP is in co-NP. 
\end{thm}

\begin{proof}
We show that the problem of deciding $s \ \not\approx^\pos \ t$
is in NP. 
To this end, we have to show that there is a certificate 
for $s \ \not\approx^\pos \ t$ that has polynomial size, and is
polynomially checkable. 
Consider the usual partition-refinement method for computing 
$\simclo$ \cite{Milner90,CONCUR98AHKV}. 
The method starts with an equivalence relation $\simeq$
that reflects propositional equivalence. 
Then, $\simeq$ is refined at most $m = |S|$ times. 
At each refinement step, some state-pairs are removed from $\simeq$.
A certificate for the removal of a pair from $\simeq$ is simply a
$\cpre$ or $\pre$ or $EU^\almost$ operator, along with a union of equivalence classes;
it is thus of size polynomial in $m$. 
Since at most $m^2$ pairs can be removed from $\simeq$, the total size
of these state-pair removal certificates is polynomial in $m$. 
This yields a polynomial-size and polynomially-checkable certificate
for $s \ \not\approx^\pos \ t$.
\end{proof}

\section{The Roles of Until and Wait-For}\label{res3}

In this section we study the roles of the until and the 
wait-for operator, and the relationship between the equivalences
induced by $\qrctl$ and $\qrctl^*$.

It is well known that in the standard branching logics $\ctl$ and $\ctl^*$,
as well as in ATL, 
the next-time operator $\nextop$ is the only temporal operator needed
for characterizing bisimulation. 
For $\qrctl$, this is not the case: the operators $\until$ and $\wait$
can increase the distinguishing power of the logics, as the following
theorem indicates.

\begin{thm}\label{thrm:diswithnext} 

The following assertions hold: 
\begin{enumerate}[\em(1)]

\item \label{part2-next} On finitely-branching MDPs, we have
  $\approx^{\pos,\nextop} \ = \ \simclo$.

\item For all MDPs, we have 
  $\approx^{\pos} \ \subseteq \ \approx^{\pos,\bigcirc}$.

\item For finite AMDPs, we have 
  $\approx^{\pos,\bigcirc} \ = \ \approx^\pos$.

\item There is a finitely-branching, infinite AMDP on which 
  $\approx^{\pos,\bigcirc} \ \not\subseteq \ \approx^{\pos}$. 

\item There is a finite, (non-alternating) MDP on which 
  $\approx^{\pos,\bigcirc} \ \not\subseteq \ \approx^{\pos}$.

\end{enumerate}
\end{thm}

\begin{proof}
{\em Assertion 1.} 
The inclusion $\approx^{\pos,\bigcirc} \ \subseteq \
\simclo$ follows from the fact that formulas 
used in the first part of the proof of 
Theorem~\ref{lemm:possub} make use only of the $\nextop$ temporal
operator, and from $\simclo \ = \ \simclo^{\nextop}$. 
To prove the inclusion $\simclo \ \subseteq \ \approx^{\pos,\bigcirc}$, 
consider two states $s, t \in S$ such that $s \not\approx^{\pos,\bigcirc} t$.
Then, there is a $\qrctl^{\pos}$ formula $\varphi$ that distinguishes them. 
From this formula we derive an $\atl$ formula $f(\varphi)$ that
also distinguishes them. 
We proceed by structural induction. 
The result is obvious for boolean operators and atomic propositions.
The cases $\phi = \exists^{\almost}{\nextop}\phi_1$
and $\phi = \exists^{\pos}{\nextop}\phi_1$ are an easy consequence of 
Lemma~\ref{lem-trans1}.

\medskip 

\noindent {\em Assertion 2.} 
Immediate, as the set of $\qrctl^\pos$ formulas without
$\until$ and $\wait$ is a subset of the set of all $\qrctl^\pos$ formulas. 

\medskip 

\noindent {\em Assertion 3.} 
The result is derived as follows: 
$
  \approx^{\pos,\bigcirc} \ \subseteq \ \simclo \ 
   = \ \approx^\pos
$.
The inclusion $\approx^{\pos,\bigcirc} \ \subseteq \ \simclo$ is a
consequence of Assertion~\ref{part2-next} of this theorem. 
The equality $\altsim \ = \ \approx^\pos$ follows by combining 
Assertion~\ref{possub-one} of Theorem~\ref{lemm:possub} 
and Assertion~\ref{simsub-one} of Theorem~\ref{lemm:simsub}. 

\medskip 

\noindent {\em Assertion 4.} 
The result follows by considering again the infinite AMDP of
Figure~\ref{fig2}.
Reasoning as in the proof of Theorem~\ref{lemm:simsub}, it holds
$(s,t) \in \ \approx^{\pos,\bigcirc}$, but $(s,t)\not \in \ \ \approx^\pos$: 
indeed, note that 
$s \models \exists^\pos (\Box q)$ and $t \not\models \exists^\pos(
\Box q)$.

\medskip 

\noindent {\em Assertion 5.} 
The result is a consequence of Theorem~\ref{lemm:simsub},
Assertion~\ref{part-counterexconc}, and of the present theorem,
Assertion~\ref{part2-next}: the same MDP used to show $\simclo \
\not\subseteq \ \approx^\pos$, depicted in Figure~\ref{fig4}, also
shows $\approx^{\pos,\bigcirc} \ \not\subseteq \ \approx^{\pos}$.
\end{proof}

\section{Linear Time Nesting}\label{res4}

The logics $\ctl$ and $\ctl^*$ induce the same equivalence, namely,
bisimulation. 
Similarly, ATL and ATL* both induce alternating bisimulation. 
We show here that $\qrctl$ and $\qrctl^*$ induce the same equivalences
on finite, alternating MDPs, but we show that for infinite, or
non-alternating, MDPs, $\qrctl^*$ induces finer relations than $\qrctl$. 
These results are summarized by the following theorem. 

\begin{thm}
\label{theo-equi-pos}
The following assertions hold: 
\begin{enumerate}[\em(1)]
\item For all MDPs, we have 
  $\approx^\pos_* \ \subs \ \approx^\pos$.
\item For all finite AMDPs, we have 
  $\approx^\pos_* \ = \ \approx^\pos$.
\item There is a finitely-branching, infinite AMDP, on which 
  $\approx^\pos \ \not\subs \ \approx^\pos_*$.
\item There is a finite MDP on which 
  $\approx^\pos \ \not\subs \ \approx^\pos_*$.
 \end{enumerate}
\end{thm}

Before presenting the proof of this result, it is useful to recall
some facts about Rabin automata, Markov decision processes, and
probabilistic verification. 

\subsubsection*{Rabin automata and temporal logic} 

An \emph{infinite-word automaton over $\AP$} is a tuple
$A=(L,L_\init,\autlab{\cdot},{\Delta})$, where $L$ is a finite set of
locations, $L_\init \subs L$ is the set of initial locations, 
$\autlab{\cdot}: L \mapsto 2^\AP$ is a labeling
function that associates with each location $l \in L$ the set
$\autlab{l} \subs \AP$ of predicates that are true at $l$, 
and ${\Delta}: L \mapsto 2^L$ is the transition relation.
 The automaton $A$ is deterministic if the following conditions hold: 
\begin{enumerate}[$\bullet$] 
\item for all $\eta \subs \AP$, there is a unique $l \in L_\init$ with
  $\autlab{l} = \eta$; 
\item for all $l \in L$ and all $\eta \subs \AP$, there is 
  $l' \in \Delta(l)$ with $\autlab{l'} = \eta$; 
\item for all $l, l', l'' \in L$, we have that $l', l'' \in \Delta(l)$
  and $l' \neq l''$ implies $\autlab{l'} \neq \autlab{l''}$.
\end{enumerate}
The set of paths of $A$ is $\pathlang(A) = \set{l_0, l_1, l_2, \ldots
  \mid l_0 \in L_\init \und \forall k \geq 0 \qdot l_{k+1} \in
  \Delta(l_k)}$. A {\em Rabin acceptance condition\/} over a set $L$
is a set of pairs $F = \set{(P_1,R_1),(P_2,R_2),\ldots,(P_m,R_m)}$
where, for $1 \leq i \leq m$, we have $P_i, R_i \subs L$.  The
acceptance condition $F$ defines a set of paths over $L$.  For a path
$\tau = s_0, s_1, s_2, \ldots \in L^\omega$, we define $\Inf(\tau)$ to
be the set of locations that occur infinitely often along $\tau$.  We
define $\pathlang(F) = \set{\tau \in L^\omega \mid \exists i \in
  [1..m] \qdot (\Inf(\tau) \cap P_i = \emptyset \und \Inf(\tau) \cap
  R_i \neq \emptyset)}$.  A {\em Rabin automaton\/} $(A,F)$ is an
infinite-word automaton $A$ with set of locations $L$, together with a
Rabin acceptance condition $F$ on $L$; we associate with it the set of
paths $\pathlang(A,F) = \pathlang(A) \inters \pathlang(F)$.

Given a set of predicates $\AP$, a {\em trace\/}
$\trace \in  (2^\AP)^\omega$ {\em over $\AP$\/} is an infinite
sequence of interpretations of $\AP$; we indicate with 
$\traces(\AP) = (2^\AP)^\omega$ the set of all traces over $\AP$. 
A Rabin automaton $(A,F)$ with $A = (L,L_\init,\autlab{\cdot},{\Delta})$
induces the set of traces 
$\traces(A, F) = \set{\autlab{l_0}, \autlab{l_1}, \autlab{l_2}, \ldots \mid
l_0, l_1, l_2, \ldots \in \pathlang(A,F)}$. 
An $\ltl$ formula $\phi$ over the set of propositions $\AP$ induces
the set of traces $\traces(\phi) \subs \traces(\AP)$, defined as
usual (see, e.g., \cite{MPvol1}).  
From~\cite{vw86} it is known that for an $\ltl$ formula
$\phi$ we can construct a deterministic Rabin automaton $(A, F)$ such
that $\traces(A, F) = \traces(\phi)$.

We can now proceed to prove Theorem~\ref{theo-equi-pos}. 

\subsubsection*{Proof of Theorem~\ref{theo-equi-pos}} 

\begin{proof}

The first assertion is obvious. 
For the other assertions, we proceed as follows. 

\noindent \emph{Assertion 2.} 
Let $G=(S,\Acts,\mov,\trans, \lab{\cdot})$ be a finite, 
alternating MDP.
Since $\qrctl$ is a fragment of $\qrctl^*$, it follows that 
$\approx^\pos_* \ \subs \ \approx^\pos$.
To prove $\approx^\pos \subs \approx^\pos_*$, we show that
if there exists a $\qrctl^*$ formula that distinguishes two states $s$
and $t$, then there also exists a $\qrctl$ formula that distinguishes
$s$ and $t$.
We focus on formulas of the type $\exists^{\pos}\varphi$ and
$\exists^{\almost}\varphi$, where $\varphi$ is an $\ltl$ formula. 
The generalization to the complete
logic follows by structural induction and duality.
Thus, assume that there are two states $s^*, t^* \in S$ and 
$\alpha \in \set{\almost,\pos}$ such that 
$s^* \sat \exists^\alpha \phi$ and $t^* \not\sat \exists^\alpha \phi$.
Let $(A,F)$ be a deterministic Rabin automaton 
such that $\traces(A, F) = \traces(\phi)$, and assume that
$A =  (L, L_\init, \autlab{\cdot}, {\Delta})$ and 
$F = \{(P_1,R_1), \ldots, (P_m,R_m)\}$.
Let $G' = G \times A =  (S',\Acts,\mov',\trans', \lab{\cdot}')$ be the
MDP resulting from forming the usual synchronous product of $G$ and
$A$. 
In detail, we have:
\begin{enumerate}[$\bullet$] 
\item $S' = \set{ (s,l) \in S \times L \mid \lab{s} = \autlab{l} }$;
\item $\mov'(s,l) = \mov(s)$ for all $(s,l) \in S'$; 
\item for all $(s_1, l_1), (s_2, l_2) \in S'$ and $a \in \Acts$, we
have $\trans'((s_1,l_1),a)(s_2,l_2) = \trans(s_1,a)(s_2)$ if $l_2 \in
\Delta(l_1)$,
and $\trans'((s_1,l_1),a)(s_2,l_2) = 0$ otherwise;
\item $\lab{(s,l)} = \autlab{l}$, for all $(s,l) \in S'$. 
\end{enumerate}
Let $F'$ be the Rabin acceptance condition of $G'$,
defined by $F' = \{(P'_1,R'_1), \ldots, (P'_m,R'_m)\}$, where each
$P_i',R_i' \subs S'$ is defined as follows: 
$P_i'=\set{(s,l)\in S' \mid l \in P_i}$ and 
$R_i'=\set{(s,l)\in S' \mid l \in R_i}$.
For every $s \in S$, denote with $l_\init(s)$ the unique
$l \in L_\init$ such that $\lab{s} = \autlab{l}$.
Using the results of \cite{luca-thesis,luca-focs98,Trading04} on
the model-checking of MDPs with respect to probabilistic
temporal-logic properties, we can construct $\mu$-calculus formulas to
distinguish $(s^*, l_\init(s^*))$ and  $(t^*, l_\init(t^*))$. 
Define, first of all, the following abbreviations: 
\begin{align*}
  \hat{\psi}^\sure & = {\bigcup_{i=1}^m} \; \nu Y .\, \mu X . 
      \Bigl[ P'_i \inters \bigl( \cpre(X) \union 
      ( R'_i \inters \cpre(Y) ) \bigr) \Bigr] \\
  \hat{\psi}^\almost & = {\bigcup_{i=1}^m} \; \nu Y \,. \mu X . 
   \Bigl[ P'_i \inters \bigl( \apre(Y,X) \union 
      ( R'_i \inters \cpre(Y) ) \bigr) \Bigr] \\
  \hat{\psi}^\nullo & = {\bigcup_{i=1}^m} \; \nu Y .\, \mu X . 
      \Bigl[ P'_i \inters \bigl( \pre(X) \union 
      ( R'_i \inters \pre(Y) ) \bigr) \Bigr]. 
\end{align*}
On the basis of the above formulas, define: 
\begin{align*}
  \psi^\sure & = \mu W \qdot \bigl( \hat{\psi}^\sure \union \cpre(W) \bigr) \\
  \psi^\almost & = \nu Z \qdot \mu W \qdot \bigl( 
      \apre(Z,W) \union \hat{\psi}^\almost \bigr) \\ 
  \psi^\pos & = \mu W \qdot \bigl( \hat{\psi}^\almost \union \pre(W) \bigr) \\
  \psi^\nullo & = \mu W \qdot \bigl( \hat{\psi}^\nullo \union \pre(W) \bigr).
\end{align*}
For $\alpha \in \set{\sure, \almost, \pos, \nullo}$ and $s \in S$, we have:
\[
  (s, l_\init(s)) \in \sem{\psi^\alpha}_{G'}
  \quad \text{\textit{iff}} \quad 
  s \sat \exists^\alpha \phi
\]
so that, in particular, 
$(s^*,l_\init(s^*)) \in \sem{\psi^\alpha}_{G'}$ and 
$(t^*,l_\init(t^*)) \not\in \sem{\psi^\alpha}_{G'}$. 
Hence, the formula $\psi^\alpha$ is a $\mu$-calculus witness, on $G'$,
of the distinction between $s^*$ and $t^*$. 
We now show how to transform $\psi^\alpha$, first into a
$\mu$-calculus formula to be evaluated on $G$, and then into a
$\qrctl$ formula to be evaluated on $G$. 
This will show that $s^* \ \not\approx^\pos \ t^*$, as required. 

To obtain a $\mu$-calculus formula on $G$, from $\psi^\alpha$
we construct a $\mu$-calculus formula $\gamma^\alpha$ with the
following property: 
for all $s \in S$, we have $s \in \sem{\gamma^\alpha}_G$ iff 
$(s, l_\init(s)) \in \sem{\psi^\alpha}_{G'}$.
The idea, taken from \cite{luca-lics01}, is as follows. 

First, $\psi^\alpha$ can be rewritten in {\em equational form\/}
\cite{CleavelandBhat96}, as a sequence of blocks 
$\block'_1,\ldots,\block'_{k}$, where $\block'_1$ is the
innermost block and $\block'_{k}$ the outermost block. 
Each block $\block'_j$, for $1 \leq j \leq k$, has the form 
$\var_j = \lambda e_j$, where $\lambda \in \set{\mu,\nu}$, and
where $e_j$ is an expression not containing $\mu$, $\nu$, in
which all the occurrences of the variables $\var_1, \ldots, \var_k$
have positive polarity \cite{CleavelandBhat96}; the output variable is
$\var_k$. 

From this formula, we obtain another formula $\gamma^\alpha$, also in
equational form, with sets of variables $\set{\var_i^l \mid 
1 \leq i \leq k \und l \in L} \union \set{\var_{k+1}}$. 
Formula $\gamma^\alpha$ simulates on $G$ the evaluation of
$\psi^\alpha$ on $G'$: for each 
variable $\var_i$, with $1 \leq i \leq k$, formula $\gamma^\alpha$ contains the
set of variables $\set{\var_i^l \mid l \in L}$, where the value of
$\var_i$ at location $l \in L$ is encoded as the value of 
$\var_i^l$ at $s$.
The formula $\psi$ consists of the blocks
$\block_1,\ldots,\block_{k}$, plus an additional block $\block_{k+1}$.  
For $1 \leq i \leq k$, the block $\block_i$ contains the equations for
the variables $\set{\var_i^l \mid l \in L}$. 
The equation for $\var_i^l$ is obtained from the equation for $\var_i$
as follows:
\begin{enumerate}[$\bullet$]
\item replace each variable $\var_i$ on the left-hand side with the
  variable $\var_i^l$; 
\item replace $P_j$ (resp.\ $R_j$), for $1 \leq j \leq m$, with
  $S$ if $l \in P_j$ (resp.\ $l \in R_j$), and with $\emptyset$ 
  if $l \not\in P_j$ (resp.\ $l \not\in R_j$); 
\item replace $\cpre(\var_h)$, for variable $1 \leq h \leq k$, with 
$\cpre(\bigcup_{l' \in \Delta(l)} \var_h^{l'})$; 
\item intersect the right-hand side with $\bigcap_{\ap \in \autlab{l}}
  \ap \inters \bigcap_{\ap \in \AP \setm \autlab{l}} \no \ap$. 
\end{enumerate}
The block $\block_{k+1}$ consists of only one equation 
$\var_{k+1} = \bigcup_{l \in L_\init} \var_k^l$, and can be either a
$\mu$ or a $\nu$-block. 
The output variable is $\var_{k+1}$. 

The result of the above transformation is a $\mu$-calculus formula
$\gamma^\alpha$ on $G$ containing only the operators $\cpre$ and
$\apre$. 
By (\ref{sem-next-sure}) and Lemma~\ref{lem-trans3}, both operators
can be encoded in $\qrctl$. 
Then, proceeding as in the first part of the proof of
Theorem~\ref{lemm:simsub}, we can ``unroll'' the computation of the
fixpoints of the $\mu$-calculus formulas, since we know that each
fixpoint converges in at most $|S|$ iterations. 
The result of these two transformations is a $\qrctl$ formula
$\lambda^\alpha$, such that $s^* \sat \lambda^\alpha$ and 
$t^* \not\sat \lambda^\alpha$, as required. 

\medskip 
\noindent \emph{Assertion 3.}
Consider the AMDP $G$ with state space 
$S = (\set{1,2,3} \times \nats) \union \set{0}$.
The only successor of $0$ is $0$ itself.
States of the type $\tuple{i,2n}$, for $i \in \set{1,2,3}$ 
(i.e., \emph{even states})
belong to player~1, while odd states belong to player~$p$.
For all $n \geq 0$ we have:
  $\Gamma(\tuple{1,2n}) = \Gamma(\tuple{3,2n})= \set{a, b}$ and
  $\Gamma(\tuple{2,2n}) = \set{a, b, c}$,
where, for all $i \in \set{1,2,3}$:
\begin{align*}
  &\dest(\tuple{i,2n}, a) = \set{\tuple{i,2n}}   \\
  &\dest(\tuple{i,2n}, b) = \set{\tuple{i,2n+1}} \\
  &\dest(\tuple{2,2n}, c) = \set{\tuple{3,2n+1}}.
\end{align*}
Player~$p$ states starting with $1$ or $2$ lead to the next state
in their chain and to the sink state $0$ with equal probability.
Formally, $\Gamma(\tuple{i,2n+1}) = \set{x}$ and
\begin{align*}
  &\trans(\tuple{1,2n+1}, x)(\tuple{1,2n+2}) = 
     \trans(\tuple{1,2n+1}, x)(0) = \\
  &\trans(\tuple{2,2n+1}, x)(\tuple{2,2n+2}) =
     \trans(\tuple{2,2n+1}, x)(0) = 1/2.
\end{align*}
Finally, states starting with $3$ obey the following distribution.
\begin{align*}
  &\trans(\tuple{3,2n+1}, x)(\tuple{2,2n+2}) = \exp(-1/2^n) \\
  &\trans(\tuple{3,2n+1}, x)(0) = 1 - \exp(-1/2^n).
\end{align*}
Observe that $G$ is a finitely-branching, infinite AMDP.
We take $\AP=\set{q}$, and we ask that the predicate $q$ be true at
all odd states. 
Then, by induction on the structure of a $\qrctl$ formula, it is not
hard to see that $\tuple{1,0} \approx^\pos \tuple{2,0}$.
On the other hand, we have $\tuple{2,0} \sat \exists^\pos \bo \diam q$ and 
$\tuple{1,0} \not\sat \exists^\pos \bo \diam q$.

\medskip

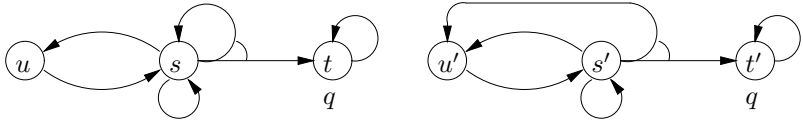
\begin{figure}[t]
\centering 
\subfigure{\setlength{\unitlength}{0.00066667in}
\begingroup\makeatletter\ifx\SetFigFont\undefined%
\gdef\SetFigFont#1#2#3#4#5{%
  \reset@font\fontsize{#1}{#2pt}%
  \fontfamily{#3}\fontseries{#4}\fontshape{#5}%
  \selectfont}%
\fi\endgroup%
{\renewcommand{\dashlinestretch}{30}
\begin{picture}(2918,925)(0,-10)
\put(1358,459){\ellipse{300}{300}}
\put(1283,369){\makebox(0,0)[lb]{{\SetFigFont{10}{12.0}{\rmdefault}{\mddefault}{\updefault}$s$}}}
\put(2558,459){\ellipse{300}{300}}
\put(2483,369){\makebox(0,0)[lb]{{\SetFigFont{10}{12.0}{\rmdefault}{\mddefault}{\updefault}$t$}}}
\put(158,459){\ellipse{300}{300}}
\put(83,369){\makebox(0,0)[lb]{{\SetFigFont{10}{12.0}{\rmdefault}{\mddefault}{\updefault}$u$}}}
\put(2733.000,634.000){\arc{353.553}{2.9997}{7.9959}}
\blacken\path(2553.248,732.602)(2558.000,609.000)(2611.973,720.297)(2553.248,732.602)
\put(1573.625,674.625){\arc{450.781}{2.8462}{8.1494}}
\blacken\path(1325.058,728.226)(1358.000,609.000)(1385.040,729.701)(1325.058,728.226)
\put(1358.000,168.375){\arc{318.750}{5.2023}{10.5056}}
\blacken\path(1532.157,235.055)(1433.000,309.000)(1485.694,197.092)(1532.157,235.055)
\put(758.000,-66.000){\arc{1500.000}{4.0689}{5.3559}}
\blacken\path(393.460,623.424)(308.000,534.000)(425.487,572.687)(393.460,623.424)
\put(758.000,984.000){\arc{1500.000}{0.9273}{2.2143}}
\blacken\path(1122.540,294.576)(1208.000,384.000)(1090.513,345.313)(1122.540,294.576)
\put(1770.500,496.500){\arc{237.171}{5.0341}{6.6049}}
\path(1508,459)(2408,459)
\blacken\path(2288.000,429.000)(2408.000,459.000)(2288.000,489.000)(2288.000,429.000)
\put(2483,84){\makebox(0,0)[lb]{{\SetFigFont{10}{12.0}{\rmdefault}{\mddefault}{\updefault}$q$}}}
\end{picture}
}}
\subfigure{\setlength{\unitlength}{0.00066667in}
\begingroup\makeatletter\ifx\SetFigFont\undefined%
\gdef\SetFigFont#1#2#3#4#5{%
  \reset@font\fontsize{#1}{#2pt}%
  \fontfamily{#3}\fontseries{#4}\fontshape{#5}%
  \selectfont}%
\fi\endgroup%
{\renewcommand{\dashlinestretch}{30}
\begin{picture}(2918,936)(0,-10)
\put(1358,459){\ellipse{300}{300}}
\put(1283,369){\makebox(0,0)[lb]{{\SetFigFont{10}{12.0}{\rmdefault}{\mddefault}{\updefault}$s'$}}}
\put(2558,459){\ellipse{300}{300}}
\put(2483,369){\makebox(0,0)[lb]{{\SetFigFont{10}{12.0}{\rmdefault}{\mddefault}{\updefault}$t'$}}}
\put(158,459){\ellipse{300}{300}}
\put(83,369){\makebox(0,0)[lb]{{\SetFigFont{10}{12.0}{\rmdefault}{\mddefault}{\updefault}$u'$}}}
\put(2733.000,634.000){\arc{353.553}{2.9997}{7.9959}}
\blacken\path(2553.248,732.602)(2558.000,609.000)(2611.973,720.297)(2553.248,732.602)
\put(1358.000,168.375){\arc{318.750}{5.2023}{10.5056}}
\blacken\path(1532.157,235.055)(1433.000,309.000)(1485.694,197.092)(1532.157,235.055)
\put(758.000,-66.000){\arc{1500.000}{4.0689}{5.3559}}
\blacken\path(393.460,623.424)(308.000,534.000)(425.487,572.687)(393.460,623.424)
\put(758.000,984.000){\arc{1500.000}{0.9273}{2.2143}}
\blacken\path(1122.540,294.576)(1208.000,384.000)(1090.513,345.313)(1122.540,294.576)
\put(1770.500,496.500){\arc{237.171}{5.0341}{6.6049}}
\put(1562.808,681.115){\arc{457.554}{4.8008}{8.0959}}
\put(282.117,784.883){\arc{253.574}{2.9360}{4.9180}}
\path(1508,459)(2408,459)
\blacken\path(2288.000,429.000)(2408.000,459.000)(2288.000,489.000)(2288.000,429.000)
\path(1583,909)(308,909)
\path(158,759)(158,609)
\blacken\path(128.000,729.000)(158.000,609.000)(188.000,729.000)(128.000,729.000)
\put(2483,84){\makebox(0,0)[lb]{{\SetFigFont{10}{12.0}{\rmdefault}{\mddefault}{\updefault}$q$}}}
\end{picture}
}}
   \caption{An MDP where $s \approx^\pos s'$ and 
     $s \not\approx^\pos_* s'$. }
   \label{fig-mdp-differentstart}
\end{figure}

\noindent \emph{Assertion 4.} 
Consider the MDP depicted in Figure~\ref{fig-mdp-differentstart}. 
By induction on the structure of a $\qrctl$ formula, it is not hard to
see that $s \approx^\pos s'$. 
On the other hand, for 
$\phi = \exists^\almost (\diam q \und \bo \exists^\pos \nextop q)$ 
we have $s \sat \phi$, $s' \not\sat \phi$. 
\end{proof}

We do not provide an algorithm for computing $\approx^\pos_*$ on
non-alternating MDPs. Identifying such an algorithm is an open problem.

\section*{Acknowledgements}

This research was supported in part by the NSF grants CCR-0132780,
CNS-0720884 and CNS-0834812, and by a BAEF grant. 

\bibliographystyle{alpha}
\bibliography{main}

\newcommand{\etalchar}[1]{$^{#1}$}
\begin{thebibliography}{dAKN{\etalchar{+}}00}

\bibitem[ABGS91]{bisim-hard91}
C.~\'Alvarez, J.~L. Balc\'azar, J.~Gabarr\'o, and M.~S\'antha.
\newblock Parallel complexity in the design and analysis of concurrent systems.
\newblock In {\em PARLE '91: Proc. on Parallel architectures and languages
  Europe}. Springer-Verlag, 1991.

\bibitem[AHK02]{ATL02}
R.~Alur, T.A. Henzinger, and O.~Kupferman.
\newblock Alternating time temporal logic.
\newblock {\em J. ACM}, 49:672--713, 2002.

\bibitem[AHKV98]{CONCUR98AHKV}
R.~Alur, T.A. Henzinger, O.~Kupferman, and M.Y. Vardi.
\newblock Alternating refinement relations.
\newblock In {\em {CONCUR 98}: Concurrency Theory. 9th Int.\ Conf.}, volume
  1466 of {\em Lect. Notes in Comp. Sci.}, pages 163--178. Springer-Verlag,
  1998.

\bibitem[ASB{\etalchar{+}}95]{BerkP95}
A.~Aziz, V.~Singhal, F.~Balarin, R.K. Brayton, and A.L.
  Sangiovanni-Vincentelli.
\newblock It usually works: The temporal logic of stochastic systems.
\newblock In {\em Computer Aided Verification}, volume 939 of {\em Lect. Notes
  in Comp. Sci.} Springer-Verlag, 1995.

\bibitem[BC96]{CleavelandBhat96}
G.~Bhat and R.~Cleaveland.
\newblock Efficient model checking via the equational $\mu$-calculus.
\newblock In {\em Proc. 11th IEEE Symp. Logic in Comp. Sci.}, pages 304--312,
  1996.

\bibitem[BdA95]{bda95}
A.~Bianco and L.~de~Alfaro.
\newblock Model checking of probabilistic and nondeterministic systems.
\newblock In {\em Found. of Software Tech. and Theor. Comp. Sci.}, volume 1026
  of {\em Lect. Notes in Comp. Sci.}, pages 499--513. Springer-Verlag, 1995.

\bibitem[Ber95]{Bertsekas95}
D.P. Bertsekas.
\newblock {\em Dynamic Programming and Optimal Control}.
\newblock Athena Scientific, 1995.
\newblock Volumes I and II.

\bibitem[BS01]{SegalaAxioms}
E.~Bandini and R.~Segala.
\newblock Axiomatizations for probabilistic bisimulation.
\newblock In {\em Proc. 28th Int. Colloq. Aut. Lang. Prog.}, volume 2076 of
  {\em Lect. Notes in Comp. Sci.}, pages 370--381. Springer-Verlag, 2001.

\bibitem[CdAH04]{Trading04}
K.~Chatterjee, L.~de~Alfaro, and T.A. Henzinger.
\newblock Trading memory for randomness.
\newblock In {\em QEST~04}. {IEEE} Computer Society Press, 2004.

\bibitem[CE81]{skeletons}
E.M. Clarke and E.A. Emerson.
\newblock Design and synthesis of synchronization skeletons using branching
  time temporal logic.
\newblock In {\em Proc. Workshop on Logic of Programs}, volume 131 of {\em
  Lect. Notes in Comp. Sci.}, pages 52--71. Springer-Verlag, 1981.

\bibitem[CY95]{CY95}
C.~Courcoubetis and M.~Yannakakis.
\newblock The complexity of probabilistic verification.
\newblock {\em J. ACM}, 42(4):857--907, 1995.

\bibitem[dA97a]{luca-thesis}
L.~de~Alfaro.
\newblock {\em Formal Verification of Probabilistic Systems}.
\newblock PhD thesis, Stanford University, 1997.
\newblock Technical Report STAN-CS-TR-98-1601.

\bibitem[dA97b]{luca-stacs97-prob}
L.~de~Alfaro.
\newblock Temporal logics for the specification of performance and reliability.
\newblock In {\em Proc. of Symp. on Theor. Asp. of Comp. Sci.}, volume 1200 of
  {\em Lect. Notes in Comp. Sci.}, pages 165--176. Springer-Verlag, 1997.

\bibitem[dAFMR05]{EMSOFT05}
L.~de~Alfaro, M.~Faella, R.~Majumdar, and V.~Raman.
\newblock Code-aware resource management.
\newblock In {\em {EMSOFT 05}: ACM Conference on Embedded Software}, Lect.
  Notes in Comp. Sci. Springer-Verlag, 2005.

\bibitem[dAH00]{luca-lics00}
L.~de~Alfaro and T.A. Henzinger.
\newblock Concurrent omega-regular games.
\newblock In {\em Proc. 15th IEEE Symp. Logic in Comp. Sci.}, pages 141--154,
  2000.

\bibitem[dAHK98]{luca-focs98}
L.~de~Alfaro, T.A. Henzinger, and O.~Kupferman.
\newblock Concurrent reachability games.
\newblock In {\em Proc. 39th IEEE Symp. Found. of Comp. Sci.}, pages 564--575.
  IEEE Computer Society Press, 1998.

\bibitem[dAHM01]{luca-lics01}
L.~de~Alfaro, T.A. Henzinger, and R.~Majumdar.
\newblock From verification to control: Dynamic programs for omega-regular
  objectives.
\newblock In {\em Proc. 16th IEEE Symp. Logic in Comp. Sci.}, pages 279--290.
  {IEEE} Press, 2001.

\bibitem[dAKN{\etalchar{+}}00]{luca-kwiat2000}
L.~de~Alfaro, M.~Kwiatkowska, G.~Norman, D.~Parker, and R.~Segala.
\newblock Symbolic model checking of concurrent probabilistic processes using
  {MTBDDs} and the {Kronecker} representation.
\newblock In {\em {TACAS}: Tools and Algorithms for the Construction and
  Analysis of Systems}, volume 1785 of {\em Lect. Notes in Comp. Sci.}, pages
  395--410. Springer-Verlag, 2000.

\bibitem[Der70]{Derman}
C.~Derman.
\newblock {\em Finite State {Markovian} Decision Processes}.
\newblock Academic Press, 1970.

\bibitem[DGJP99]{DGJP99}
J.~Desharnais, V.~Gupta, R.~Jagadeesan, and P.~Panangaden.
\newblock Metrics for labelled markov systems.
\newblock In {\em {CONCUR'99}: Concurrency Theory. 10th Int.\ Conf.}, volume
  1664 of {\em Lect. Notes in Comp. Sci.}, pages 258--273. Springer, 1999.

\bibitem[HJ94]{HanJon94}
H.~Hansson and B.~Jonsson.
\newblock A logic for reasoning about time and probability.
\newblock {\em Formal Aspects of Computing}, 6(5):512--535, 1994.

\bibitem[KNP00]{KNP_PRISM00}
M.~Kwiatkowska, G.~Norman, and D.~Parker.
\newblock Verifying randomized distributed algorithms with prism.
\newblock In {\em Workshop on Advances in Verification (WAVE'00)}, 2000.

\bibitem[Koz83]{Kozen83mu}
D.~Kozen.
\newblock Results on the propositional $\mu$-calculus.
\newblock {\em Theoretical Computer Science}, 27(3):333--354, 1983.

\bibitem[Mil90]{Milner90}
R.~Milner.
\newblock Operational and algebraic semantics of concurrent processes.
\newblock In J.~van Leeuwen, editor, {\em Handbook of Theoretical Computer
  Science}, volume~B, pages 1202--1242. Elsevier Science Publishers
  (North-Holland), Amsterdam, 1990.

\bibitem[MP91]{MPvol1}
Z.~Manna and A.~Pnueli.
\newblock {\em The Temporal Logic of Reactive and Concurrent Systems:
  Specification}.
\newblock Springer-Verlag, New York, 1991.

\bibitem[PSL00]{PSL00}
A.~Pogosyants, R.~Segala, and N.~Lynch.
\newblock Verification of the randomized consensus algorithm of {A}spnes and
  {H}erlihy: a case study.
\newblock {\em Distributed Computing}, 13(3):155--186, July 2000.

\bibitem[Seg95]{SegalaT}
R.~Segala.
\newblock {\em Modeling and Verification of Randomized Distributed Real-Time
  Systems}.
\newblock PhD thesis, MIT, 1995.
\newblock Technical Report MIT/LCS/TR-676.

\bibitem[SL94]{SL94}
R.~Segala and N.A. Lynch.
\newblock Probabilistic simulations for probabilistic processes.
\newblock In {\em {CONCUR'94}: Concurrency Theory. 5th Int.\ Conf.}, volume 836
  of {\em Lect. Notes in Comp. Sci.}, pages 481--496. Springer-Verlag, 1994.

\bibitem[ST05]{SegalaTurrini05}
R.~Segala and A.~Turrini.
\newblock Comparative analysis of bisimulation relations on alternating and
  non-alternating probabilistic models.
\newblock In {\em QEST~05}. {IEEE}, 2005.

\bibitem[Sto02]{Sto02b}
M.I.A. Stoelinga.
\newblock Fun with {FireWire}: Experiments with verifying the {IEEE1394} root
  contention protocol.
\newblock In {\em Formal Aspects of Computing}, 2002.

\bibitem[VW86]{vw86}
M.Y. Vardi and P.~Wolper.
\newblock Automata theoretic techniques for modal logics of programs.
\newblock {\em J. Comp. Sys. Sci.}, 32:183--221, 1986.

\end{thebibliography}

\end{document}